\documentclass[showpacs,aps,prd,reprint,superscriptaddress,nofootinbib,longbibliography,preprintnumbers]{revtex4-1}
\usepackage[colorlinks=true, pdfstartview=FitV, linkcolor=magenta,citecolor=blue, urlcolor=magenta,
bookmarks=true, bookmarksnumbered=true, breaklinks]{hyperref}
\usepackage{graphicx}

\usepackage{here}
\usepackage{url}
\usepackage{ulem}
\usepackage{amsmath,amssymb,bm,color,longtable,mathrsfs,amsfonts,slashed}

\newcommand{\Slash}[1]{{\ooalign{\hfil/\hfil\crcr$#1$}}}

\begin{document}
\begin{flushright}
\end{flushright}

\title{Two-pion exchange potential in $D^{(*)}D^{(*)}$ system}

\author{Daiki Suenaga}
\affiliation{Kobayashi-Maskawa Institute for the Origin of Particles and the Universe, Nagoya University, Nagoya 464-8602, Japan}
\affiliation{Research Center for Nuclear Physics, Osaka University, Ibaraki 567-0048, Japan}

\author{Masayasu Harada}
\affiliation{Kobayashi-Maskawa Institute for the Origin of Particles and the Universe, Nagoya University, Nagoya 464-8602, Japan}
\affiliation{
    Department of Physics, Nagoya University, Nagoya 464-8602, Japan
}
\affiliation{Advanced Science Research Center, Japan Atomic Energy Agency, Tokai 319-1195, Japan}

\date{\today}

\begin{abstract}
Motivated by the recent HAL QCD lattice results on the $D$ - $D^*$ potential at nearly physical point $m_\pi=146.4$ MeV, we theoretically examine effects from the two-pion exchange in (semi-)long-range parts of the $DD^*$ system. We employ the framework of heavy-meson chiral perturbation theory with heavy-quark spin symmetry. The pion exchanges up to next-to-next-to-leading order (N$^2$LO) are taken into account. In order to separate the (semi-)long-range contributions, we apply the dispersion-relation method to one-loop diagrams. As a result, we find that the lattice data on the $D$ - $D^*$ potential in the regime of $0.5\, {\rm fm} \lesssim r$ is reproduced by adjusting the unknown ${\cal O}(p^2)$ couplings. In particular, it turns out that contributions from isospin-independent N$^2$LO triangle diagrams play a central role in reproducing the lattice data of the form $V^{DD^*}\sim \left({\rm e}^{-m_\pi r}/r\right)^2$ in $1.0\, {\rm fm} \lesssim r\lesssim 2.0\, {\rm fm}$. We also present predictions of other $D^{(*)}$ - $D^{(*)}$ potentials, where the similar two-pion exchange tail is predicted. Our findings provide useful information on the (semi-)long-range regime of the $D$ - $D^*$ potential focusing on two-pion exchanges.
 \end{abstract}


\maketitle

\section{Introduction}
\label{sec:Introduction}

Investigation of exotic hadrons which cannot be classified by the conventional quark model is one of the central subjects in hadron physics. Among them, doubly-charm tetraquark, i.e., $T_{cc}^+$, reported by the LHCb collaboration in 2021 has been attracting much attention~\cite{LHCb:2021vvq}, since this state possesses two $c$ quarks and in this sense is regarded as a genuine exotic state. Besides, its mass lies below $DD^*$ threshold by only several keV, which implies that $T_{cc}^+$ state could be dominated by the so-called molecule state of $D$ and $D^*$ mesons, although other possibilities, i.e., a compact state, a double diquark configuration, and kinetic effects such as the triangle singularity, and so on, are being considered [see, e.g., Ref.~\cite{Chen:2022asf} for a review and references therein]. 

Once the hadronic molecule picture is adopted, exotic states are basically treated as a quantum mechanical two-body problem with a potential, and thus it is important to gain understandings into the hadron-hadron potential properly. In particular, when the binding energy is considerably small like $T_{cc}^+$ state, precise delineation of long-range parts of the potential is necessary~\cite{Yamaguchi:2019vea,Meng:2022ozq}.

Hadron-hadron interactions are in principle derived from the microscopic dynamics of the quantum chromodynamics (QCD).
Thus, the first-principles lattice simulation based on QCD is a powerful tool so as to gain deeper insights into inter-hadron interactions, and the HAL QCD collaboration has shown simulation results on various hadron-hadron potentials~\cite{Ishii:2006ec,Ishii:2012ssm,Aoki:2020bew}. In Ref.~\cite{Lyu:2023xro}, utilizing the lattice simulation with $m_\pi = 146.4$ MeV, the HAL QCD collaboration reported that in a semi-long range of $1.0\, {\rm fm} \lesssim r \lesssim 2.0\, {\rm fm}$ in a local $D$ - $D^*$ potential (${}^3S_1;I=0$ channel), there appears a two-pion exchange (TPE) tail whose shape is $V^{DD^*}\sim \alpha\left({\rm e}^{-m_\pi r}/r\right)^2$ with $\alpha=-43.4\, {\rm MeV}\,  {\rm fm}^2$, while theoretically the one-pion exchange (OPE) is expected to asymptotically survive. This behavior is one of the noteworthy results found by the HAL QCD collaboration, since when we regard TPE as an $S$-wave sigma-meson exchange, the tail would take the simple Yukawa-type form of $\sim {\rm e}^{-m_\sigma r}/r$ with $m_\sigma \approx 2m_\pi$.

Motivated by the above peculiar TPE tail, in this work we theoretically examine the TPE as well as the OPE contributions in the (semi-)long-range regime of $D$ - $D^*$ potential. Pions are regarded as the Nambu-Goldstone (NG) bosons associated with the spontaneous breakdown of chiral symmetry, and accordingly, interactions among pions, $D$ and $D^*$ mesons are described by the chiral dynamics. We employ the heavy-meson chiral perturbation theory (HMChPT) constructed upon a systematic low-energy expansion, where heavy-quark spin symmetry (HQSS) is also taken into account~\cite{Falk:1991nq,Neubert:1993mb,manohar2000heavy}.

The TPE effects based on the ChPT framework were investigated in various systems, not only in $N$ - $N$ but also in two-body systems including heavy flavors~\cite{Bernard:1996gq,Kaiser:1997mw,Epelbaum:2003gr,Epelbaum:2003xx,Epelbaum:2004fk,Liu:2012vd,Epelbaum:2014efa,Epelbaum:2014sza,Brambilla:2015rqa,Xu:2017tsr,TarrusCastella:2018php,Wang:2018atz,Wang:2019nvm,Xiao:2020ozd,Wang:2019ato,Meng:2019ilv,Wang:2021kos,Lu:2021gsb,Xu:2021vsi,Wang:2022jop,Zhai:2023ejo}. In particular, investigations of nucleon-pion systems have been intensively conducted with abundant experimental data over the past decades, including more than three nucleon forces. Recent advances of ChPT studies concerning nucleons are summarized in, e.g., Refs.~\cite{Epelbaum:2019kcf,Machleidt:2024bwl}.

Evaluation of the TPEs is slightly complicated since basically we need to evaluate the corresponding one-loop diagrams, unlike the simple OPEs. One concise and promising way to derive the (semi)long-range potential mediated by the TPEs is the dispersion-relation method~\cite{Kaiser:1997mw,Epelbaum:2003gr,Brambilla:2015rqa,TarrusCastella:2018php}. Within this treatment, short-range contributions which may contain various ambiguities from messy quark dynamics are successfully separated. Hence we adopt this dispersion-relation method to gain insights into the (semi)long-range TPE effects in the $D$ - $D^*$ potential. We note that while effects from other light-flavored mesons, e.g., $\sigma$, $\rho$, $\omega$ meson exchanges are phenomenologically
expected, we do not explicitly incorporate them by the virtue of systematic low-energy description of HMChPT where such resonances are dynamically generated by multi pions, in this exploratory work.

The TPE as well as OPE effects in $P^{(*)} P^{(*)}$ ($P=D,\bar{B}$) system within the ChPT framework were studied in detail in Ref~\cite{Xu:2017tsr,Wang:2018atz,Xu:2021vsi,Wang:2022jop,Zhai:2023ejo,Xu:2025xrl} to gain deeper understandings into the formation of bound and resonance states. Among them, Ref.~\cite{Wang:2022jop} was concerned with the TPEs at next-to-leading order (NLO) and at next-to-next-to-leading order (N$^2$LO) in addition to the leading order (LO) OPE, based on the HMChPT, with which the resonance saturation model~\cite{Bernard:1996gq} was adopted to fix the unknown low-energy constants. Although the tail of $V^{DD^*}\sim \alpha\left({\rm e}^{-m_\pi r}/r\right)^2$ was successfully obtained by N$^2$LO TPEs in this reference, the authors concluded that this behavior is not dominant in the range of $1.0\, {\rm fm} \lesssim r \lesssim 2.0\, {\rm fm}$. Meanwhile, in Ref.~\cite{Zhai:2023ejo} it was claimed that the lattice data~\cite{Lyu:2023xro} can be reasonably reproduced by only LO OPE and NLO TPE contributions based on the relativistically covariantized HMChPT, but with an unphysical value of the pion decay constant $f_\pi$. 

In this paper, we present another concise description of the TPE tail of $V^{DD^*}\sim \alpha\left({\rm e}^{-m_\pi r}/r\right)^2$ in $1.0\, {\rm fm} \lesssim r \lesssim 2.0\, {\rm fm}$; we show that this lattice result can be reproduced with no contact interactions and form factors, based on the simplest HMChPT framework with large N$^2$LO contributions. As for the consistency check, we also present predictions of other $D^{(*)}D^{(*)}$ potential with the same HMChPT set up, which would be checked by future lattice simulations. 
We emphasize that this work only aims to shed light on the pion-exchange nature at (semi-)long regime in $D^{(*)}D^{(*)}$ potentials, so that no dedicated treatment of the short-range interactions are addressed. Therefore, we do not discuss possibility of the bound or resonance state where all the range of potential subtly plays an important role.

This paper is organized as follows. In Sec.~\ref{sec:Model}, we introduce our HMChPT Lagrangian which describe interactions among pions, $D$ and $D^*$ mesons consistently with chiral symmetry. 
Based on it, in Sec.~\ref{sec:Amplitude} we derive a $D$ - $D^*$ potential including OPE and TPE diagrams, with the help of dispersion-relation method. Then, in Sec.~\ref{sec:PotentialNumerical} we present numerical results of $D$ - $D^*$ potential and compare with the HAL QCD data. Section~\ref{sec:OtherChannels} is devoted to predictions of the other $S$-wave $D^{(*)}D^{(*)}$ potentials. Finally, we conclude the present work in Sec.~\ref{sec:Conclusions}.

\section{Model}
\label{sec:Model}


In order to study the local potential generated by pion exchanges in between $D$ and $D^*$ mesons, we adopt the HMChPT framework~\cite{manohar2000heavy}. This effective theory is based on the HQSS of heavy quarks and chiral symmetry for light quarks, which allows us to describe a systematic expansion of interactions among the heavy mesons and pions. In this section, we introduce this HMChPT Lagrangian up to ${\cal O}(p^2)$ coupled to pions with HQSS preserved. We note that contact interactions are not incorporated since our main aim is to delineate the pion exchange effects in the (semi-)lone-range regime.

Within the HMChPT treatment, $0^-$ and $1^-$ heavy mesons ($D_v$ and $D_v^*$) are treated as a doublet owing to the HQSS as 
\begin{eqnarray}
H = \frac{1+\Slash{v}}{2}(i\gamma_5D_v+\Slash{D}_v^*) \ ,
 \end{eqnarray}
where $v^\mu$ stands for a velocity of the doublet and in what follows we will choose the rest frame: $v^{\mu} = (1,{\bm 0})$. This heavy-meson doublet obeys transformation properties of $H \to S_h H$ and $H \to Hh^\dagger$, with $S_h \in SU(2)_{\rm HQS}$ and $h \in SU(2)_{L+R}$. 
Meanwhile, pion fields are always accompanied by derivatives due to its nonlinear realization. In particular, it is convenient to utilize the so-called Maurer-Cartan $1$-forms~\cite{Bando:1987br}
\begin{eqnarray}
\alpha_{\perp}^\mu &=& \frac{1}{2i}(\partial^\mu\xi_R\xi_R^\dagger - \partial^\mu\xi_L\xi_L^\dagger)\ , \nonumber\\
\alpha_{\parallel}^\mu &=& \frac{1}{2i}(\partial^\mu\xi_R\xi_R^\dagger +\partial^\mu\xi_L\xi_L^\dagger)\ , 
\end{eqnarray}
with $\xi_R=\xi_L^\dagger = {\rm exp}(i\pi^a\tau^a/2f_\pi)$ ($\tau^a$ is the Pauli matrix). Since $\xi_R\to h\,\xi_R g_R^\dagger$ and $\xi_{L}\to h\xi_{L} g_L^\dagger$ under $SU(2)_L\times SU(2)_R$ chiral transformation with $g_{R(L)} \in SU(2)_{R(L)}$, one can easily show that the $1$-forms exhibit the following chiral transformation laws:
\begin{eqnarray}
\alpha_{\perp}^\mu \to h\alpha_{\perp}^\mu h^\dagger\ , \ \ \alpha_{\parallel}^\mu \to h\alpha_{\parallel}^\mu h^\dagger-i\partial^\mu hh^\dagger\ .
\end{eqnarray}

Thus, an HMChPT Lagrangian up to ${\cal O}(p^2)$ preserving $SU(2)$ HQSS and $SU(2)_L\times SU(2)_R$ chiral symmetry is derived to be
\begin{eqnarray}
{\cal L}_{\rm HMChPT} =  {\cal L}_{{\cal O}(p^1)} +   {\cal L}_{{\cal O}(p^2)}\ ,
\end{eqnarray}
where
\begin{eqnarray}
 {\cal L}_{{\cal O}(p^1)} &=&  -{\rm tr}[H(iv\cdot D)\bar{H}]_D \nonumber\\
 && + f_\pi^2\big({\rm tr}[\alpha_{\perp\mu}\alpha^\mu_\perp]_f +{\rm tr}[\hat{\chi} + \hat{\chi}^\dagger]_f\big) \nonumber\\
 &&+  g{\rm  tr}_D[H\gamma_5\gamma^\mu\alpha_{\perp,\mu}\bar{H}]_D \ , \label{LOp1}
\end{eqnarray}
and
\begin{eqnarray}
{\cal L}_{{\cal O}(p^2)} &=&  h_1 {\rm tr}[Hg_{\mu\nu}\alpha_\perp^\mu\alpha^\nu_{\perp}\bar{H}]_D + h_2 {\rm tr}[Hv_\mu v_\nu\alpha^\mu_{\perp}\alpha^\nu_{\perp}\bar{H}]_D \nonumber\\
&& + h_3 ({\rm tr}[H\gamma_\mu v_\nu \alpha^\mu_{\perp}\alpha^\nu_{\perp}\bar{H}]_D + {\rm H.c.}) \nonumber\\
&& + h_4 {\rm tr}[H(\hat{\chi}^\dagger + \hat{\chi})\bar{H}]_D \nonumber\\
&& +  i h_5 {\rm tr}[H\sigma_{\mu\nu}\alpha_\perp^\mu\alpha^\nu_{\perp}\bar{H}]_D + \cdots\ .
 \label{LOp2}
\end{eqnarray}
In these equations, the covariant derivative is defined by $D_\mu \bar{H} = (\partial_\mu-i\alpha_{\parallel,\mu})\bar{H}$, and ${\rm Tr}[]_f$ and ${\rm Tr}[]_D$ stand for trace operators for flavor and Dirac indices, respectively. The scalar spurion field transforms as $\hat{\chi}\to h\hat{\chi}h^\dagger$ from which $\hat{\chi} =\xi_L^\dagger \chi \xi_R $ with $\chi\to g_L\chi g_R^\dagger$ under chiral symmetry. This field is introduced to incorporate the pion mass systematically, thus in the end $\chi$ is replaced by $m_\pi^2/4$. The dots in Eq.~(\ref{LOp2}) stand for ${\cal O}(p^2)$ terms with derivatives acting on heavy mesons which do not contribute to the local $D$ - $D^*$ potential. It should be noted that the interactions are also constructed upon both parity and time-reversal symmetries. We again emphasize that contact interactions are neglected so as to focus on (semi-)long-range pion exchanges.


\begin{figure*}[htbp]
\centering
\hspace*{-0.7cm} 
\includegraphics[width=19cm]{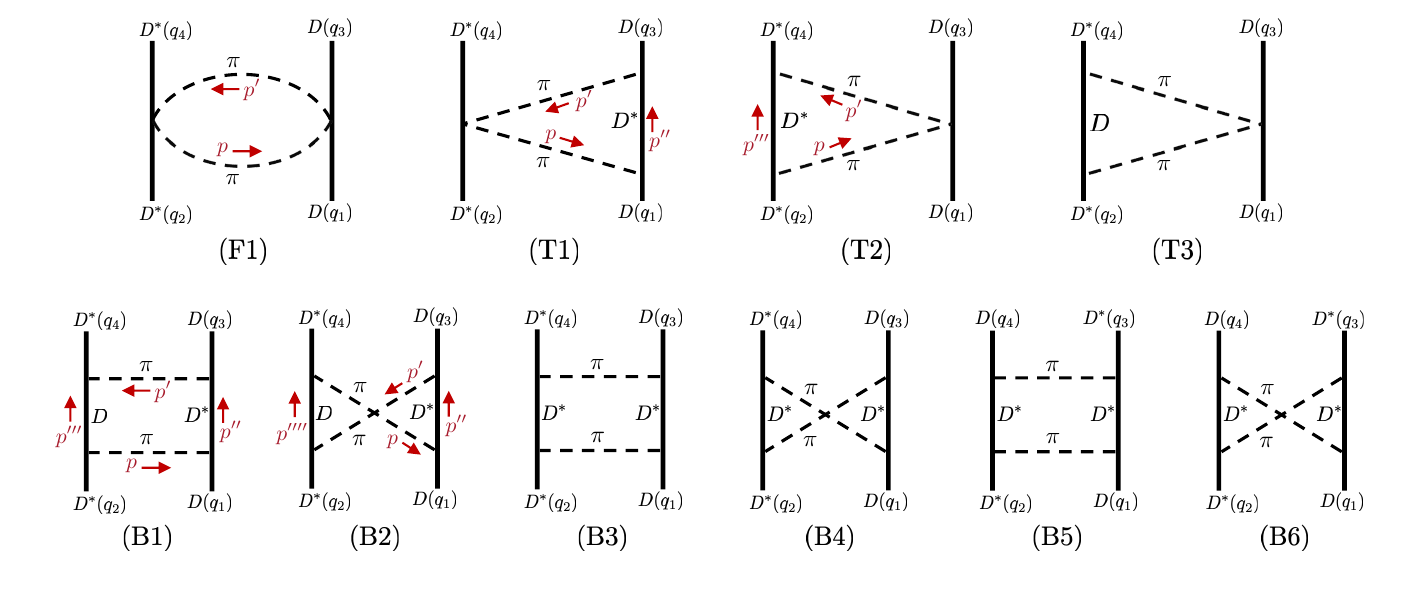}
\caption{NLO [${\cal O}(\nu^2)$] $DD^*\to DD^*$ amplitude generated by TPEs. There are football diagram (F1), triangle diagrams (T1) - (T3), and planar and nonplanar box diagrams (B1) - (B6). Here, $p'=p+q$, $p''=p+q_1$, $p'''=q_2-p$ and $p''''=q_4+p$ ($q=q_1-q_3=q_4-q_2$), which are used in the explicit computations in Appendix~\ref{sec:OneLoops}.}
\label{fig:DDst_NLO}
\end{figure*}   

\begin{figure}[htbp]
\centering
\hspace*{-0.5cm} 
\includegraphics[width=8cm]{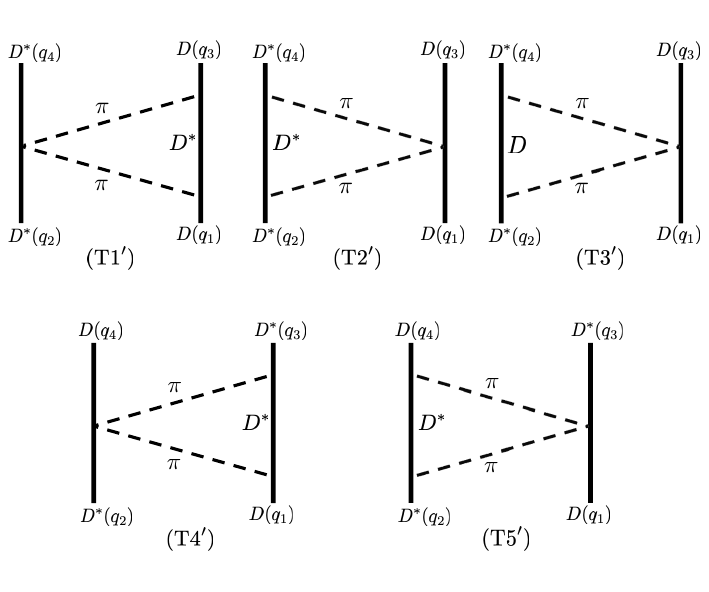}
\caption{N$^2$LO [${\cal O}(\nu^3)$] $DD^*\to DD^*$ amplitude generated by TPEs. There are five types of triangle diagrams when the HQSS is imposed.}
\label{fig:DDst_NNLO}
\end{figure}   

\section{$D$ - $D^*$ Potential}
\label{sec:Amplitude}

A nonrelativistic potential in between $D$ and $D^*$ mesons is derived by the Fourier transformation of the corresponding scattering amplitude, where the heavy-mass limit is appropriately taken. Within the present HMChPT formalism, the limit can be simply reached by taking zero-energy limits of the external $D$ and $D^*$ mesons. 

Here, we define the chiral dimension $\nu = 2L + \sum_iV_i\Delta_i$ with $\Delta_i = d_i + n_i/2-2$, where $L$ and $V_i$ stand for the number of loops and of vertices of type $i$, respectively. In $\Delta_i$, $d_i$ and $n_i$ represent the number of derivatives or pion masses and of heavy-meson fields involved in the vertex, respectively~\cite{Xu:2017tsr}. The power counting is applied with increasing of the power of $\nu$.

\subsection{The OPE potential}
\label{sec:OPE}

At LO of ${\cal O}(\nu^0)$, the $D$ - $D^*$ potential is simply given by an OPE from two $D^*D\pi$ interactions of  ${\cal O}(p^1)$ in Eq.~(\ref{LOp1}). The corresponding $u$-channel $DD^*\to DD^*$ amplitude takes the form of
\begin{eqnarray}
{\cal M}_{\rm LO}^{DD^*} &=& (-)^{I+1}\frac{{\cal C}_Ig^2}{3f_\pi^2} \nonumber\\
&&\times\left(({\bm \epsilon}^*\cdot{\bm \epsilon})-\frac{({\bm \epsilon}^*\cdot{\bm \epsilon})m_\pi^2-S_{12}({\bm q})}{{\bm q}^2+m_\pi^2}\right)\ ,
\label{OPEAmplitude}
\end{eqnarray}
with a momentum transfer ${\bm q}$.\footnote{In what follows, the energy transfer $q_0$ will be always taken to be zero ($q_0=0$) to derive the non-relativistic potentials.} The coefficient ${\cal C}_I$ is the isospin factor of $\tau^a\otimes \tau^a$:
\begin{eqnarray}
{\cal C}_I = \left\{
\begin{array}{cc}
  -3 & (I=0) \\
  +1 & (I=1) \\
  \end{array}
 \right.\ , \label{CiDef}
\end{eqnarray}
where $I=0$ and $I=1$ correspond to the iso-singlet and iso-triplet channels, respectively. After the Fourier transformation, the first term in the parenthesis is translated into the delta function $\delta^3({\bm r})$ in the coordinate space. In the following analysis we will omit this contribution to focus on the (semi-)long-range regime of $D$ - $D^*$ potential. In Eq.~(\ref{OPEAmplitude}), ${\bm \epsilon}$ and ${\bm \epsilon}^*$ are polarization vectors of initial- and final-state $D^*$ mesons, respectively. $S_{12}({\bm q}) = 3({\bm \epsilon}^*\cdot {\bm q})({\bm \epsilon}\cdot{\bm q})-({\bm \epsilon}^*\cdot{\bm \epsilon}){\bm q}^2$ is an operator created by two rank-$2$ tensor operators: $[{\bm \epsilon}^*\otimes{\bm \epsilon}]_2$ and $[{\bm q}\otimes {\bm q}]_2$, and thus this operator drives the mixing between $S$-wave  and $D$-wave $DD^*$ states.  

The iso-singlet and iso-triplet wave functions, $|DD^*\rangle_{I=0}$, $|D^*D\rangle_{I=0}$, and $|DD^*\rangle_{I=1}$, $|D^*D\rangle_{I=1}$, are antisymmetric and symmetric with respect to light-flavor interchanges, respectively. Besides, for the $S$-wave $DD^*$ state spin and angular parts are symmetric. Thus, since the $DD^*$ system is bosonic whose total wave function must be symmetric, 
$|DD^*({}^3S_1;I=0)\rangle$ and $|DD^*({}^3S_1;I=1)\rangle$ are constructed in an antisymmetric and symmetric way with respect to the interchange of $D$ and $D^*$:
\begin{eqnarray}
|DD^* ({}^3S_1;I=0)\rangle &=& \frac{1}{\sqrt{2}}\Big(|DD^*\rangle_{I=0}-|D^*D\rangle_{I=0} \Big) \ , \nonumber\\
|DD^* ({}^3S_1; I=1)\rangle &=& \frac{1}{\sqrt{2}}\Big(|DD^*\rangle_{I=1} + |D^*D\rangle_{I=1} \Big)\ . \nonumber\\ \label{DDstWF}
\end{eqnarray}
The overall sign factor $(-)^{I+1}$ in Eq.~(\ref{OPEAmplitude}) emerges due to this (anti)symmetrization since the amplitude is $u$-channel.

 In the present analysis, we neglect the $S$ - $D$ mixing as suggested by the HAL QCD data~\cite{Lyu:2023xro} and concentrate on pure $S$-wave $D$ - $D^*$ potential. Since the $S$-wave $DD^*$ state carries no angular momentum while $S_{12}({\bm q})$ contains the rank-2 $[{\bm q}\otimes {\bm q}]_2$ piece, $S_{12}({\bm q})$ sandwiched by two $S$-wave $DD^*$ state vanishes. Besides, for $S$-wave case the inner product of $({\bm \epsilon}^*\cdot{\bm \epsilon})$ is trivially reduced to $1$. Those reduction enables us to straightforwardly derive the following OPE potential from Eq.~(\ref{OPEAmplitude}):
\begin{eqnarray}
V_{\rm LO}^{DD^*} = (-)^{I+1}{\cal N}\frac{{\cal C}_Ig^2m_\pi^2}{12\pi f_\pi^2}\frac{{\rm e}^{-m_\pi r}}{r}\ . \label{VLO}
\end{eqnarray}
The additional factor ${\cal N}$ represents a normalization factor connecting our HMChPT formalism to a nonrelativistic potential, i.e., ${\cal N} = 1/4$. Equation~(\ref{VLO}) indicates that the OPE generates a Yukawa-type repulsive potential at least at long range.

\subsection{The TPE potential}
\label{sec:TPE}

The HAL QCD results indicate the TPE tail in a semi-long range of $1.0\, {\rm fm}\lesssim r\lesssim 2.0\,  {\rm fm}$ in the $D$ - $D^*$ potential, the behavior of which is $ V^{DD^*}\sim\alpha \left({\rm e}^{-m_\pi r}/r\right)^2$ ($\alpha=-43.4\, {\rm MeV}\,  {\rm fm}^2$). Thus, here we evaluate $DD^*\to DD^*$ scattering amplitude mediated by two pions to gain insights into this notable TPE tail.

While the OPE in $DD^*\to DD^*$ scattering amplitude is derived at LO of ${\cal O}(\nu^0)$, TPE contributions appear at NLO of ${\cal O}(\nu^2)$, which are derived from combinations of the ${\cal O}(p^1)$ vertices in Eq.~(\ref{LOp1}). We collectively depict the diagrams in Fig.~\ref{fig:DDst_NLO}. There are one football diagram (F1), three triangle diagrams (T1) - (T3), and six planar and nonplanar box diagrams (B1) - (B6). In addition, we take into account the N$^2$LO diagrams whose chiral dimension is of ${\cal O}(\nu^3)$, which can be generated by mixed couplings from ${\cal O}(p^1)$ and ${\cal O}(p^2)$ interactions, Eqs.~(\ref{LOp1}) and~(\ref{LOp2}). Those diagrams are depicted in Fig.~\ref{fig:DDst_NNLO}.

 In the present work, first we derive the respective imaginary parts, and then the real parts are evaluated via the dispersion relation
\begin{eqnarray}
{\cal M} (s) =   \frac{1}{\pi}\int_{4m_\pi^2}^\infty ds'\frac{{\rm Im}{\cal M}(s')}{s'-s-i\epsilon}\ , \label{Dispersion}
\end{eqnarray}
where $s=q^2=-{\bm q}^2$ with a momentum transfer $q=q_1-q_3=q_4-q_2$. Computation of imaginary parts of the amplitudes in Fig.~\ref{fig:DDst_NLO} is lengthly, so we leave it in Appendix~\ref{sec:OneLoops}. As summarized in this Appendix, imaginary parts of the NLO and N$^2$LO iso-singlet $DD^*\to DD^*$ amplitude mediated by TPEs are evaluated to be
\begin{eqnarray}
 {\rm Im}[{\cal M}^{DD^*}_{\rm NLO}] &=& \frac{{\cal C}_I}{192\pi f_\pi^4}  \Big[-(s-4m_\pi^2)+2g^2(5s-8m_\pi^2) \nonumber\\
 && +g^4(23s-20m_\pi^2)\Big] \sqrt{1-\frac{4m_\pi^2}{s}} \nonumber\\
&&  +\frac{(-)^Ig^4}{16\pi f_\pi^4}s \sqrt{1-\frac{4m_\pi^2}{s}} \ , \label{ImNLO}
\end{eqnarray}
and
\begin{eqnarray}
&& {\rm Im}[{\cal M}^{DD^*}_{\rm N^2LO}] \nonumber\\
&=& \frac{3g^2}{64f_\pi^4}\frac{h_1(s-2m_\pi^2)^2 + 2h_4m_\pi^2(s-2m_\pi^2)}{\sqrt{s}} ({\bm \epsilon}^*\cdot{\bm \epsilon}) \, , \nonumber\\ 
&& + \frac{(-)^{I+1}{\cal C}_Ih_5g^2}{192 f_\pi^4}\frac{s-4m_\pi^2}{\sqrt{s}} \Big[2s ({\bm \epsilon}^*\cdot{\bm \epsilon}) +S_{12}({\bm q})\Big]  \ , \nonumber\\
 \label{ImNNLO}
\end{eqnarray}
respectively.

We note that the box diagrams would contain 2-particle reducible (2PR) diagrams which lead to the pinch singularity pointed by Weinberg~\cite{Weinberg:1990rz,Weinberg:1991um}. Those contributions break the power counting of HMChPT and thus must be removed. Our present loop calculations illustrated in Appendix~\ref{sec:OneLoops} are essentially equivalent to the subtraction done in Ref.~\cite{Xu:2017tsr,Xu:2025xrl}.

Since the potential is obtained from the Fourier transformation of the corresponding amplitude, the dispersion relation~(\ref{Dispersion}) enables us to derive it by the following simple formula~\cite{Kaiser:1997mw,Epelbaum:2003gr,Brambilla:2015rqa,TarrusCastella:2018php}
\begin{eqnarray}
V(r) &=& {\cal N} \int \frac{d^3q}{(2\pi)^3}\left[- {\cal M}(s)\right]{\rm e}^{i{\bm p}\cdot {\bm r}} \nonumber\\
&=& - \frac{{\cal N}}{\pi}\int_{4m_\pi^2}^\infty ds'\int \frac{d^3q}{(2\pi)^3} \frac{{\rm Im}{\cal M}(s')}{s'+{\bm q}^2-i\epsilon} {\rm e}^{i{\bm p}\cdot {\bm r}} \nonumber\\ 
&=& - \frac{{\cal N}}{2\pi^2 r}\int_{2m_\pi}^\infty d\mu\, \mu\, {\rm e}^{-\mu r}{\rm Im}{\cal M}(\mu^2)\ , \label{Potential}
\end{eqnarray}
where 
the relation $s' = \mu^2$ has been used in the second line to arrive at the last expression. One-loop amplitudes potentially contain ultraviolet divergences; the identity~(\ref{Dispersion}) may not hold owing to non-negligible contributions from $|s'|\to\infty$, and accordingly we need to adopt the subtracted version of the dispersion relation~(\ref{Dispersion}). This subtraction generates additional terms to the potential~(\ref{Potential}) which depend on subtraction scheme, i.e., renormalization scheme or condition. Those cumbersome contributions are, however, translated into the delta function $\delta^3(r)$ or its derivatives in terms of the potential in coordinate space, which only reflect very short-range information between $D$ and $D^*$. Since the present work aims to delineate (semi-)long-range parts of the potential derived from TPEs, we neglect such arbitrary contributions and adopt the formula~(\ref{Potential}) to {\it uniquely} determine the potential.\footnote{When taking short-range dynamics from the TPEs more seriously, one may employ the (twice-)subtracted dispersion relation~\cite{Reinert:2017usi,Wang:2022jop}. }

With the help of the formula~(\ref{Potential}), Eqs.~(\ref{ImNLO}) and~(\ref{ImNNLO}) lead to the following potentials of NLO and N$^2$LO:
\begin{eqnarray}
 V^{DD^*}_{\rm NLO} &=& \frac{{\cal C}_Im_\pi}{128 \pi^3 f_\pi^4 r^4}\Big[\Big(1 - 10 g^2-23g^4\Big)x K_0(2x) \nonumber\\
&&+  \Big(1- 10g^2 - 23g^4 - 4\big(g^2  + 3g^4\big)x^2\Big)  K_1(2x)\Big]
\nonumber\\
&&+  \frac{(-)^{I+1}m_\pi}{128 \pi^3 f_\pi^4 r^4}\Big[12g^4 x K_0(2x) \nonumber\\
&& + \big(12g^4 +8g^4x^2\big)  K_1(2x)\Big] \  , 
  \label{VNLO}
\end{eqnarray}
and
\begin{eqnarray}
V^{DD^*}_{\rm N^2LO} &=&-\frac{3g^2}{128\pi^2 f_\pi^4 r^6}{\rm e}^{-2x}\Big[  h_1 (6+12x+10x^2  +4x^3 \nonumber\\
&&  + x^4) + h_4x^2(1+x)^2 \Big] + \frac{(-)^{I}{\cal C}_Ig^2h_5}{96\pi^2 f_\pi^4 r^6}{\rm e}^{-2x}\times \nonumber\\
&&(3+6x+5x^2+2x^3) \ , 
 \label{VNNLO} 
\end{eqnarray}
respectively, with $x=m_\pi r$. It is worth emphasizing that in the long-range limit $r\gg m_\pi^{-1}$, the potential asymptotically behaves as
\begin{eqnarray}
V^{DD^*}_{\rm NLO} \sim \frac{g^2\left[-{\cal C}_I+\big(2(-)^{I+ 1}-3{\cal C}_I\big)g^2\right]m_\pi^{5/2}}{64\pi^{5/2}f_\pi^4}  \frac{{\rm e}^{-2m_\pi r}}{r^{5/2}}\ , \nonumber\\
\label{VNLOAs}
\end{eqnarray}
and
\begin{eqnarray}
V^{DD^*}_{\rm N^2LO} \sim -\frac{3g^2(h_1+h_4)m_\pi^4}{128\pi^3 f_\pi^4}\frac{{\rm e}^{-2m_\pi r}}{r^2}\ , \label{VNNLOAs}
\end{eqnarray}
respectively. In other words, the behavior of $V^{DD^*} \sim \alpha\left({\rm e}^{-m_\pi r}/r\right)^2$ is achieved from N$^2$LO TPEs but not from NLO~\cite{Wang:2022jop}.

\begin{figure}[htbp]
\centering
\hspace*{-0.5cm} 
\includegraphics[width=8cm]{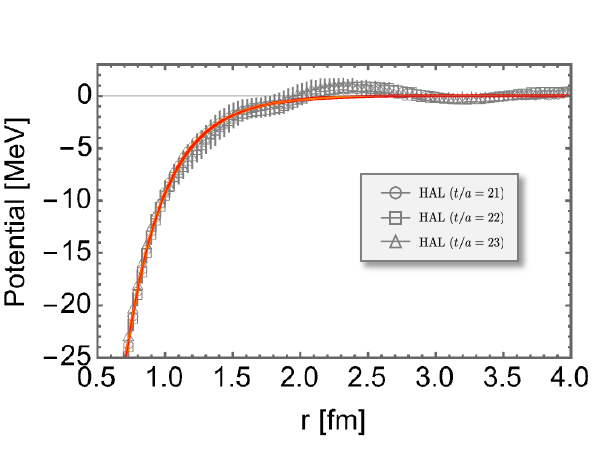}
\caption{Fitted $D$ - $D^*$ potential with the HAL QCD results~\cite{Lyu:2023xro}. }
\label{fig:Pot_DDst}
\end{figure}   

\begin{table}[t]
\begin{center}
  \begin{tabular}{cccccc}  \hline\hline
$f_\pi$ & $m_\pi $ & $g$ & $h_1$ & $h_4$ & $h_5$ \\
\, [MeV] & [MeV] & & [MeV$^{-1}$] & [MeV$^{-1}$] & [MeV$^{-1}$] \\ \hline
$93$ & $146.4$ & $0.55$ & $0.328$ & $0.005$  & $-0.491$ \\
\hline \hline
 \end{tabular}
\caption{
Values of parameters. Here, $f_\pi$ and $g$ are determined by physical values, while $m_\pi$ is fixed by the HAL QCD data. The remaining $h_1$, $h_4$ and $h_5$ are determined to fit the HAL QCD potential in Fig.~\ref{fig:Pot_DDst}.
}
\label{tab:Parameters}
\end{center}
\end{table}

\begin{figure}[htbp]
\centering
\hspace*{-0.7cm} 
\includegraphics[width=8cm]{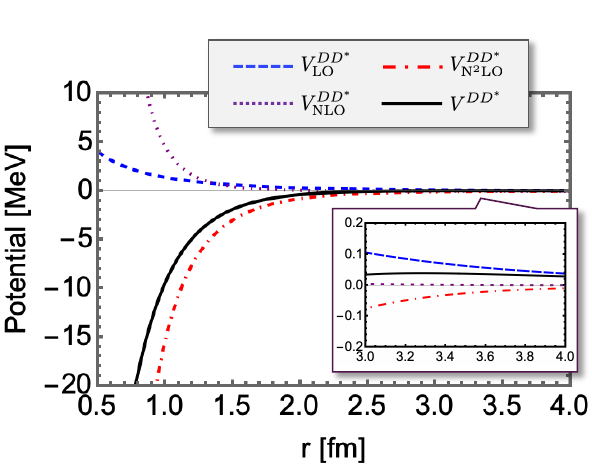}
\caption{$D$ - $D^*$ potential at LO (blue dashed), NLO (purple dotted), and N$^2$LO (red dash-dotted), with the parameters in Table~\ref{tab:Parameters}.
The black solid curve is the sum of three contributions~(\ref{VTot}). }
\label{fig:PotDetail_DDst}
\end{figure}   

\begin{figure}[htbp]
\centering
\hspace*{-0.7cm} 
\includegraphics[width=8cm]{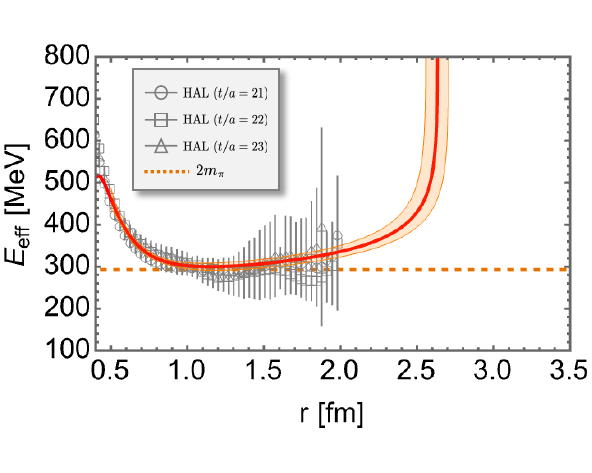}
\caption{$r$ dependency of the effective mass defined in Eq.~(\ref{EeffDef}) with the HAL QCD data~\cite{Lyu:2023xro}. The orange band corresponds to the allowed region within 95\% confidence level in fitting the potential in Fig.~\ref{fig:Pot_DDst}.}
\label{fig:Eeff_DDst}
\end{figure}   

\begin{figure*}[htbp]
\centering
\hspace*{-0.7cm} 
\includegraphics[width=18cm]{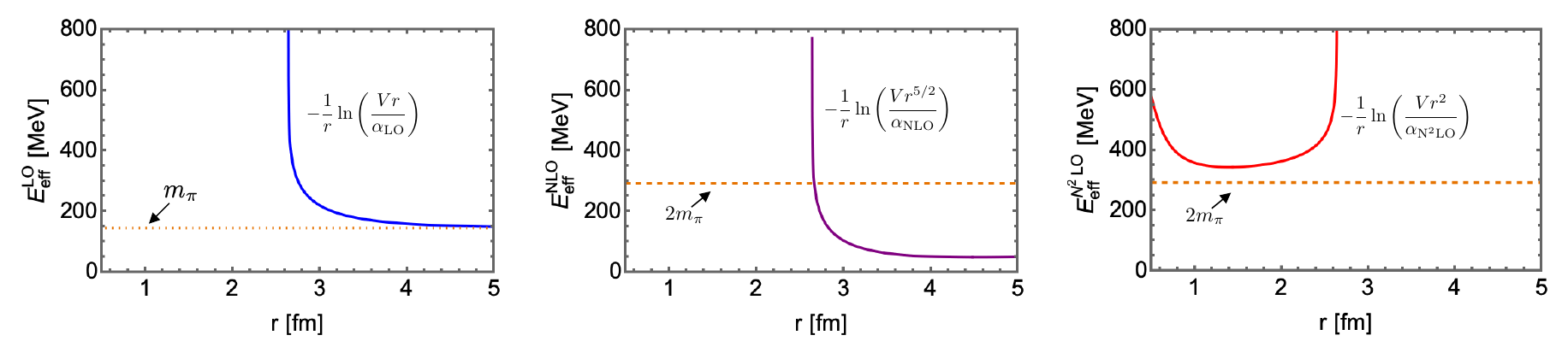}
\caption{$r$ dependencies of $E_{\rm eff}^{\rm LO}$, and $E_{\rm eff}^{\rm NLO}$ and $E_{\rm eff}^{\rm N^2LO}$ defined in Eq.~(\ref{EeffEach}). }
\label{fig:Dominance}
\end{figure*}   

\section{Comparison with the lattice data}
\label{sec:PotentialNumerical}

In this section, we show numerical results on the $D$ - $D^*$ potential to make comparisons with the HAL QCD data. Here, we employ the pion mass of $m_\pi=146.4$ MeV to compare with the HAL QCD results~\cite{Lyu:2023xro} more directly. This pion mass is already close to the physical point $m_\pi\simeq138$ MeV, so that we may be allowed to employ physical values for the $D^*D\pi$ coupling and pion decay constant, $g=0.55$~\cite{CLEO:2001sxb}
and $f_\pi=93$ MeV~\cite{ParticleDataGroup:2024cfk}. We determine the unknown ${\cal O}(p^2)$ couplings, $h_1$, $h_4$ and $h_5$, by fitting them to the HAL QCD data.

Depicted in Fig.~\ref{fig:Pot_DDst} is the resultant $D$ - $D^*$ potential
\begin{eqnarray}
V^{DD^*} = V_{\rm LO}^{DD^*} + V_{\rm NLO}^{DD^*} + V_{\rm N^2LO}^{DD^*}\ ,\label{VTot}
\end{eqnarray} 
with the lattice data~\cite{Lyu:2023xro}. We have utilized the $\chi^2$ fitting in a range of $0.524\, {\rm fm} \leq r \leq 1.98\, {\rm fm}$ with $t/a=21$ ($a = 0.0846$ fm is the lattice spacing) to get the red curve,\footnote{We have fitted the HAL QCD results with $t/a=21$ as a representative, since the other results with $t/a=22$ and $23$ are quantitatively similar.} with which the fitted ${\cal O}(p^2)$ parameters are given by
\begin{eqnarray}
h_1 &=& 0.328\, {\rm MeV}^{-1}\ , \nonumber\\
h_4 &=& 0.005\, {\rm MeV}^{-1}\ ,  \nonumber\\
h_5 &=& -0.491\, {\rm MeV}^{-1}\ .  \label{FittedHs}
\end{eqnarray}
The normalized $\chi^2$ is $\chi^2/({\rm d.o.f.}) =0.034$. The orange band represents a region allowed within $95\%$ confidence level around the best-fitted curve. The figure implies that the $D$ - $D^*$ potential at the (semi-)long range is fitted satisfactorily up to the N$^2$LO calculation. It should be noted that the oscillating behavior for $2.0\, {\rm fm} \lesssim r$ observed by the HAL QCD, which could be artifact of the lattice analysis, is not reproduced. For readability purposes, we tabulate the parameters with the fitted ones~(\ref{FittedHs}) in Table~\ref{tab:Parameters}.

It should be noted that when fitting the potential in a range of $1.0\, {\rm fm}\lesssim r \lesssim 2.0\, {\rm fm}$, the allowed region within 95 \% confidence level is significantly broadened for $r \lesssim 1.0\, {\rm fm}$ with fitted parameters  quantitatively similar to Eq.~(\ref{FittedHs}). The detailed analysis in the short-range part is beyond the scope of the present work. 

In order to take a closer look at each contribution, in Fig~\ref{fig:PotDetail_DDst}, we separately depict $V_{\rm LO}^{DD^*}$ (blue dashed), $V_{\rm NLO}^{DD^*}$ (purple dotted), and $V_{\rm N^2LO}^{DD^*}$ (red dash-dotted), with the parameters in Table~\ref{tab:Parameters}. The black solid curve denotes the sum of the three contributions~(\ref{VTot}). The figure indicates that the LO OPE and NLO TPE potentials are repulsive, while the N$^2$LO TPE is attractive with the present parameters.\footnote{The repulsive NLO TPE was also discussed in Refs.~\cite{Xu:2017tsr,Xu:2025xrl}.} Hence, the attractive shape of HAL QCD result is mainly reproduced by the N$^2$LO contributions. We note that when $r\gg m_\pi^{-1}$, the hierarchy becomes $|V_{\rm LO}^{DD^*}| > |V_{\rm N^2LO}^{DD^*}| > |V_{\rm NLO}^{DD^*}| $ reflecting the asymptotic powers of $V_{\rm LO}^{DD^*} \sim {\rm e}^{-m_\pi r}/r$, $V_{\rm NLO}^{DD^*} \sim {\rm e}^{-2m_\pi r}/r^{5/2}$ and $V_{\rm N^2LO}^{DD^*} \sim {\rm e}^{-2m_\pi r}/r^2$ [see Eqs.~(\ref{VLO}),~(\ref{VNLOAs}) and~(\ref{VNNLOAs})].

The HAL QCD collaboration reported lattice data on 
\begin{eqnarray}
E_{\rm eff} \equiv -\frac{1}{r}\ln\left(\frac{V^{DD^*}r^2}{\alpha}\right) \label{EeffDef}
\end{eqnarray}
in addition to the potential itself, so as to quantify the ``effective mass'' in the exponent of $V^{DD^*}\sim \alpha \frac{{\rm e}^{-E_{\rm eff}r}}{r^2}$ intensively, and they indeed confirmed that $E_{\rm eff} \sim 2m_\pi$ within error bars in the semi-long range~\cite{Lyu:2023xro}. Then, as a further consistency check, in Fig.~\ref{fig:Eeff_DDst} we present our model evaluation of $E_{\rm eff}$ together with the HAL QCD data. In this figure the value of $\alpha$ has been chosen to be 
\begin{eqnarray}
\alpha=-43.4\, {\rm MeV}\,  {\rm fm}^2 =-0.0011\,  {\rm MeV}^{-1} \label{AlphaLattice}
\end{eqnarray}
as determined by the HAL QCD collaboration. The red curve is drawn with the fitted parameters~(\ref{FittedHs}), while the orange band stands for the allowed region within $95\%$ confidence level, similarly to in Fig.~\ref{fig:Pot_DDst}.

Figure~\ref{fig:Eeff_DDst} indicates that our model result is in a good agreement with the lattice one consistently with the potential in Fig.~\ref{fig:Pot_DDst}. 
From the above analyses, the importance of sufficiently attractive N$^2$LO TPE contributions in $D$ - $D^*$ potential in the semi-long range is evident. 

The most dominant powers of LO OPE, NLO TPE and N$^2$LO TPE at large $r$ are $V_{\rm LO}^{DD^*} \sim {\rm e}^{-m_\pi r}/r$, $V_{\rm NLO}^{DD^*} \sim {\rm e}^{-2m_\pi r}/r^{5/2}$ and $V_{\rm N^2LO}^{DD^*} \sim {\rm e}^{-2m_\pi r}/r^2$, as shown in Eqs.~(\ref{VLO}), (\ref{VNLOAs}) and (\ref{VNNLOAs}), respectively.
In order to check manifestation of those dominances in the potential, we define
\begin{eqnarray}
E^{\rm LO}_{\rm eff} &=& -\frac{1}{r}\ln\left(\frac{V^{DD^*}r}{\alpha_{\rm LO}}\right)\ , \nonumber\\
E^{\rm NLO}_{\rm eff} &=& -\frac{1}{r}\ln\left(\frac{V^{DD^*}r^{5/2}}{\alpha_{\rm NLO}}\right)\ , \nonumber\\
E^{\rm N^2LO}_{\rm eff} &=& -\frac{1}{r}\ln\left(\frac{V^{DD^*}r^2}{\alpha_{\rm N^2LO}}\right)\ ,  \label{EeffEach}
\end{eqnarray}
where the coefficients $\alpha$ are the respective asymptotic factors given in Eqs.~(\ref{VLO}),~(\ref{VNLOAs}) and~(\ref{VNNLOAs}):
\begin{eqnarray}
\alpha_{\rm LO} &=& \frac{g^2m_\pi^2}{16\pi f_\pi^2} = 0.015 \ , \nonumber\\
\alpha_{\rm NLO} &=&  \frac{g^2(3+7g^2)m_\pi^{5/2}}{64\pi^{5/2}f_\pi^4} = 4.8\times 10^{-6}\, {\rm MeV}^{-3/2}\ , \nonumber\\
\alpha_{\rm N^2LO} &=& -\frac{3g^2(h_1+h_4)m_\pi^4}{128\pi^3 f_\pi^4} =  -0.0015\, {\rm MeV}^{-1}\ . \label{AlphaEach}
\end{eqnarray}
The potential $V^{DD^*}$ in Eq.~(\ref{EeffEach}) is the full result evaluated with the parameter set in Table~\ref{tab:Parameters}. Hence, if there is a regime where $V^{DD^*}$ is dominated by the LO OPE, then $E_{\rm eff}^{\rm LO}$ is close to $m_\pi$: $E_{\rm eff}^{\rm LO} \sim m_\pi$. Similarly, if $V$ is dominated by the NLO TPE or N$^2$LO TPE, one can confirm $E_{\rm eff}^{\rm NLO}\sim 2m_\pi$ or $E_{\rm eff}^{\rm N^2LO}\sim 2m_\pi$.

Depicted in Fig.~\ref{fig:Dominance} show $r$ dependencies of $E_{\rm eff}^{\rm LO}$, $E_{\rm eff}^{\rm NLO}$ and $E_{\rm eff}^{\rm N^2LO}$ defined in Eq.~(\ref{EeffEach}). First, one can clearly see $E^{\rm LO}_{\rm eff}\sim m_\pi$ for $3.0\, {\rm fm} \lesssim r$ from the left panel, which implies that the LO OPE dominates the potential in this long-range regime. Next, the right panel indicates that $E_{\rm eff}^{\rm N^2LO}\sim 2m_\pi$ for $1.0\, {\rm fm} \lesssim r\lesssim 2.0\, {\rm fm}$, although quantitatively the convergence is slightly poor. Thus, the N$^2$LO TPE dominance indeed appears in this semi-long-range regime. Meanwhile, from the middle panel, no $E_{\rm eff}^{\rm NLO}\sim  2m_\pi$ region is found for any $r$, meaning that there is no clear NLO TPE dominant range in the potential $V^{DD^*}$. Those observations are consistent with each potential explicitly delineated in Fig.~\ref{fig:PotDetail_DDst}; for $r\lesssim 2.0\, {\rm fm}$ the N$^2$LO is the most sizable, while for $3.0\, {\rm fm} \lesssim r$ the LO OPE dominates. In the middle range of $2.0\, {\rm fm} \lesssim r \lesssim 3.0\, {\rm fm}$, the three contributions comparably exist. We note that the model value of $\alpha_{\rm N^2LO}$ in Eq.~(\ref{AlphaEach}) is close to the lattice value~(\ref{AlphaLattice}).

\begin{figure*}[htbp]
\centering
\hspace*{-0.7cm} 
\includegraphics[width=18cm]{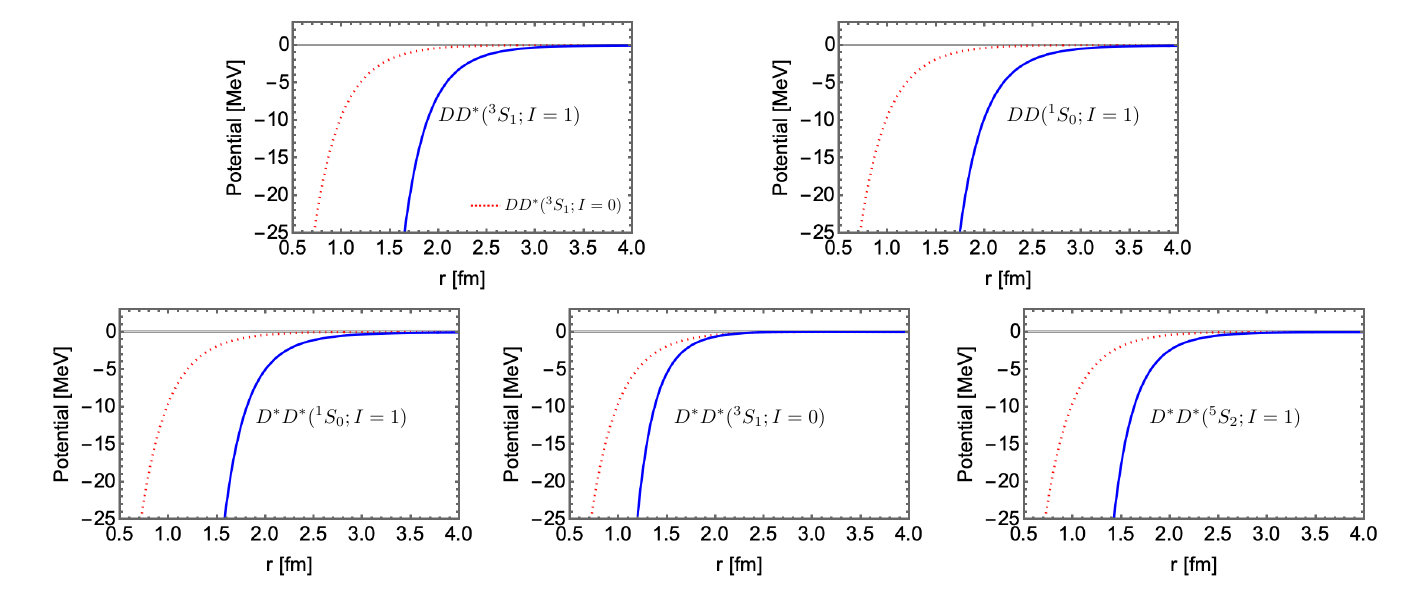}
\caption{Potential shapes for $DD^*({}^3S_1;I=1)$, $DD({}^1S_0;I=1)$, $D^*D^*({}^1S_0;I=1)$, $D^*D^*({}^3S_1;I=0)$ and $D^*D^*({}^5S_2;I=1)$ channels predicted with the parameter set in Table.~\ref{tab:Parameters}. The dotted red curve is the potential for $DD^*({}^3S_1;I=0)$ channel shown as a reference.}
\label{fig:OtherPotential}
\end{figure*}   

\begin{figure*}[htbp]
\centering
\hspace*{-0.7cm} 
\includegraphics[width=18cm]{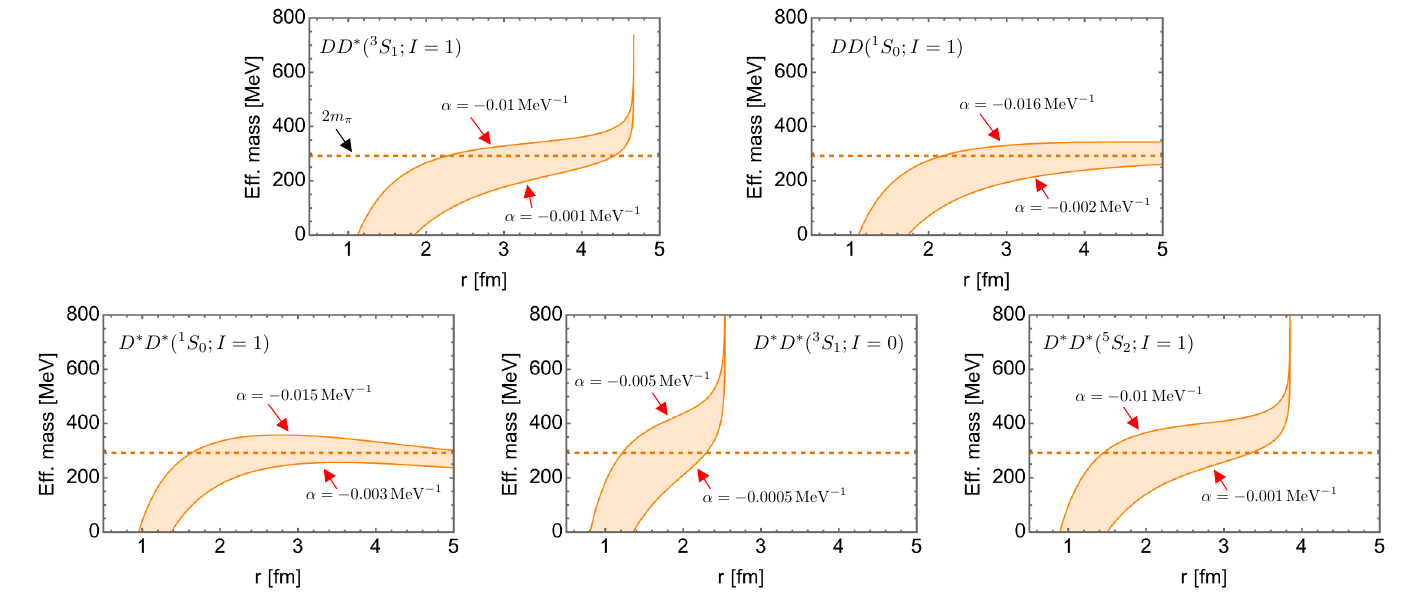}
\caption{Effective mass $E_{\rm eff} = -\frac{1}{r}\ln\left(V r^2/\alpha\right)$ for $DD^*({}^3S_1;I=1)$, $DD({}^1S_0;I=1)$, $D^*D^*({}^1S_0;I=1)$, $D^*D^*({}^3S_1;I=0)$ and $D^*D^*({}^5S_2;I=1)$ channels with the parameter set in Table.~\ref{tab:Parameters}. The orange bands are obtained by varying the values of $\alpha$ for each channel, as tabulated in Table~\ref{tab:AlphaEach}.}
\label{fig:OtherEeff}
\end{figure*}   

\section{Other channels}
\label{sec:OtherChannels}

In Sec.~\ref{sec:PotentialNumerical} we have concentrated on the $D$ - $D^*$ potential (${}^3S_1;I=0$ channel) to make a comparison with the HAL QCD data, where the TPE tail of the form $V^{DD^*}\sim \alpha \left({\rm e}^{-m_\pi r}/r\right)^2$ for $1.0\, {\rm fm} \lesssim r \lesssim 2.0\, {\rm fm}$ found on the lattice has been fitted satisfactorily up to the N$^2$LO calculations. In this section, we further study other channels of $D^{(*)}D^{(*)}$ system by employing the same parameter sets so as to provide predictions for future lattice simulations.

Here, we present the potentials for the following $S$-wave states which would be comparably doable on the lattice simulations:
\begin{eqnarray}
 DD^*({}^3S_1;I=1)\ &,& \ \ DD({}^1S_0;I=1)\ , \nonumber\\
 D^*D^*({}^1S_0; I=1)\ &,& \ \ D^*D^*({}^3S_1; I=0)\ , \nonumber\\
 D^*D^*({}^5S_2; I=1)\  .&& 
\end{eqnarray}
The respective isospin has been fixed from the symmetric nature of a bosonic system. The potentials for $DD$ and $D^*D^*$ systems are derived in Appendix~\ref{sec:DDOneLoops} and Appendix~\ref{sec:DstDstOneLoops}, which read
\begin{eqnarray}
 V^{DD}_{\rm LO} = 0\ , \label{VDDLO}
\end{eqnarray}
\begin{eqnarray}
 V^{DD}_{\rm NLO} &=& \frac{{\cal C}_Im_\pi}{128\pi^3 f_\pi^4 r^4}\Big[(1-10g^2-23g^4)xK_0(2x) + \nonumber\\
&&   \Big(1-10g^2-23g^4  - 4g^2(1+3g^2)x^2\Big)  K_1(2x)\Big] \ , \nonumber\\
\end{eqnarray}
\begin{eqnarray}
V^{DD}_{\rm N^2LO} &=&  -\frac{3g^2}{128\pi^2 f_\pi^4 r^6}{\rm e}^{-2x}\Big[  h_1 (6+12x+10x^2\nonumber\\
&&  +4x^3 + x^4) + h_4x^2(1+x)^2 \Big] \ ,  \label{VDDNNLO}
\end{eqnarray}
and
\begin{eqnarray}
V_{\rm LO}^{D^*D^*} &=& \frac{{\cal C}_I{\cal C}_Jg^2m_\pi^2}{48\pi f_\pi^2}\frac{{\rm e}^{-x}}{r}\ , \label{VLODstDst}
\end{eqnarray}
\begin{eqnarray}
 V^{D^*D^*}_{\rm NLO} &=& \frac{{\cal C}_Im_\pi}{128 \pi^3 f_\pi^4 r^4}\Big[\Big(1 - 10 g^2-23g^4\Big) x K_0(2x) +\nonumber\\
&& \Big(1- 10g^2 - 23g^4 - 4\big(g^2  + 3g^4\big)x^2\Big)  K_1(2x)\Big]
\nonumber\\
&& + \frac{{\cal C}_Jg^4m_\pi}{32 \pi^3 f_\pi^4 r^4}\Big[3xK_0(2x)  + (3 +2x^2)  K_1(2x)\Big] \ , \nonumber\\
\end{eqnarray}
\begin{eqnarray}
V^{D^*D^*}_{\rm N^2LO} &=&  -\frac{g^2}{128\pi^2 f_\pi^4 r^6}{\rm e}^{-2x}\Big[  h_1 (6+12x+10x^2\nonumber\\
&& +4x^3 + x^4) + h_4x^2(1+x)^2 \Big]  \nonumber\\
&& - \frac{{\cal C}_I{\cal C}_Jg^2h_5}{384\pi^2 f_\pi^4 r^6}{\rm e}^{-2x}\left(3+6x+5x^2+2x^3\right)  \ . \nonumber\\ \label{VDstDstNLO}
\end{eqnarray}
In these equations,  the coefficient ${\cal C}_J$ for $D^*D^*$ state is defined such that
\begin{eqnarray}
{\cal C}_J = \left\{
\begin{array}{cc}
-2 & ({\rm for}\ {}^1S_0) \\
-1 & ({\rm for}\ {}^3S_1) \\
+1 & ({\rm for}\ {}^5S_2) \\
\end{array}
\right. \ .
\end{eqnarray}
The isospin factor ${\cal C}_I$ is defined in the same way as in Eq.~(\ref{CiDef}).

Depicted in Fig.~\ref{fig:OtherPotential} are the predicted potential shapes for $DD^*({}^3S_1;I=1)$, $DD({}^1S_0;I=1)$, $D^*D^*({}^1S_0;I=1)$, $D^*D^*({}^3S_1;I=0)$ and $DD^*({}^5S_2;I=1)$ channels, with the parameter set in Table~\ref{tab:Parameters}. The dotted red curve corresponds to the potential for $DD^*({}^3S_1;I=0)$ with the same parameters. The figure indicates that all the other channels have deeper attractive potentials at (semi-)long range. 
This is mainly due to the more enhanced attractive N$^2$LO effects in these channels. Besides, the repulsive LO and NLO contributions are weakened or turn into attractive compared to the $DD^*({}^3S_1;I=0)$ channel. Here we again emphasize that our present study is devoted to delineation of the potentials at (semi-)long range from the TPE aspects, whereas the detailed prediction at short distance is beyond the scope of this work.

The HAL QCD results in Ref.~\cite{Ikeda:2013vwa} indicate that potentials for $DD({}^1S_0;I=1)$ and $DD^*({}^3S_1;I=1)$ positively evolve toward short distance for $1.0\, {\rm fm} \lesssim r$, with comparably large pion masses: $m_\pi \approx 410, 570,700$ MeV. With such heavy pion masses, the TPE contributions would be suppressed and short-range correlations that are not incorporated here could manifest themselves. Hence, we expect future lattice simulations with a more realistic pion mass.

From the analytic forms in Eqs.~(\ref{VDDNNLO}) and~(\ref{VDstDstNLO}), we can see that the asymptotic behaviors of the N$^2$LO potential universally read
\begin{eqnarray}
V_{\rm N^2LO} \sim -\frac{3g^2(h_1+h_4)m_\pi^4}{128\pi^3 f_\pi^4}\frac{{\rm e}^{-2m_\pi r}}{r^2}\ ,
\end{eqnarray}
for all the $DD^*$, $DD$ and $D^*D^*$ channels. This fact implies that in the other channels as well, the TPE tail of the form $V \sim\alpha\frac{{\rm e}^{-2m_\pi r}}{r^2}$ could be manifestly confirmed, similarly to the $DD^*({}^3S_1;I=0)$ case. In order to check this, we extend the effective mass $E_{\rm eff} = -\frac{1}{r}\ln\left(V r^2/\alpha\right)$ with $V=V_{\rm LO} + V_{\rm NLO} + V_{\rm N^2LO}$ for the other channels [see Eq.~(\ref{EeffDef})]. Since the parameter $\alpha$ is arbitrary, we will vary $\alpha$ within a certain range such that the behavior of $E_{\rm eff}\sim 2m_\pi$ is seen. Depicted in Fig.~\ref{fig:OtherEeff} are the resultant $r$ dependencies of $E_{\rm eff}$ for $DD^*({}^3S_1;I=1)$, $DD({}^1S_0;I=1)$, $D^*D^*({}^1S_0;I=1)$, $D^*D^*({}^3S_1;I=0)$ and $D^*D^*({}^5S_2;I=1)$. The orange bands correspond to the respective ranges of $\alpha$ tabulated in Table~\ref{tab:AlphaEach}.

\begin{table}[t]
\begin{center}
  \begin{tabular}{c|c}  \hline\hline
  Channel & Range of $\alpha$ [MeV$^{-1}$]\\  \hline
$DD^*({}^3S_1;I=1)$ & $-0.001 \leq \alpha \leq -0.01$ \\ 
$DD({}^1S_0;I=1)$ & $ -0.002 \leq \alpha \leq -0.016$ \\ 
$D^*D^*({}^1S_0;I=1)$ & $ -0.003\leq \alpha \leq -0.015$ \\ 
$D^*D^*({}^3S_1;I=0)$ & $-0.0005 \leq \alpha \leq -0.005$ \\ 
$D^*D^*({}^5S_2;I=1)$ & $-0.001 \leq \alpha \leq -0.01$ \\ \hline
$DD^*({}^3S_1;I=0)$ & $\alpha=-0.0011$ (from HAL) \\ 
\hline \hline
 \end{tabular}
\caption{Range of $\alpha$ for each channel. As a reference, $\alpha$ value for $DD^*({}^3S_1;I=0)$ obtained from the HAL QCD in Eq.~(\ref{AlphaLattice}) is also shown. }
\label{tab:AlphaEach}
\end{center}
\end{table}


Figure~\ref{fig:OtherEeff} demonstrates that the noteworthy behaviors of $E_{\rm eff} \sim 2m_\pi$ manifest themselves in $DD^*({}^3S_1;I=1)$, $DD({}^1S_0;I=1)$, $D^*D^*({}^1S_0;I=1)$ and $D^*D^*({}^5S_2;I=1)$ with the appropriate values of $\alpha$. For $D^*D^*({}^3S_1;I=0)$, meanwhile, the plateau is not so clear with any $\alpha$. The corresponding ranges of $r$ are dependent on the channels, as demonstrated in Fig.~\ref{fig:OtherEeff}, but are typically longer than $1.0\, {\rm fm} \lesssim r \lesssim 2.0\, {\rm fm}$, i.e., that for $DD^*({}^3S_1;I=0)$. 
This pushed-out plateau is due to the sizable attractive contaminations from TPE effects of the form $\sim \beta{\rm e}^{-2m_\pi r}/r^p$ with $p>2$. These contaminations lead to the following behavior of $E_{\rm eff}$ at smaller $r$:
\begin{eqnarray}
E_{\rm eff} &\sim& -\frac{1}{r}\ln\left(\frac{\alpha\frac{{\rm e}^{-2m_\pi r}}{r^2}+\beta\frac{{\rm e}^{-2m_\pi r}}{r^{p}}}{\alpha}r^2\right) \nonumber\\
&=& 2m_\pi - \frac{1}{r}\ln\left(1+\frac{\beta}{\alpha}\frac{1}{r^{p-2}}\right)\ .
\label{eq:E_eff cont}
\end{eqnarray}
Hence, when $\beta<0$ with $\alpha<0$ and $\frac{\beta}{\alpha}\frac{1}{r^{p-2}} \gg 1$, this $E_{\rm eff}$ is reduced to
\begin{eqnarray}
E_{\rm eff} \sim  2m_\pi + \frac{p-2}{r}\ln\left[\left(\frac{\alpha}{\beta}\right)^{\frac{1}{p-2}}r\right] \ ,
\label{eq:E_eff red}
\end{eqnarray}
illustrating that $E_{\rm eff}$ evolves from below with increasing $r$, as confirmed in Fig.~\ref{fig:OtherEeff}. For $DD^*({}^3S_1;I=0)$ channel, the relatively large repulsive NLO contributions diminish the attractive contaminations, which results in that $E_{\rm eff}$ decreases from above for $0.5\, {\rm fm}\lesssim r$ as seen in Fig.~\ref{fig:Eeff_DDst}.

The behavior of $E_{\rm eff}$ at larger $r$ is controlled by the fate of LO OPE. When the LO OPE is repulsive, $E_{\rm eff}$ positively diverges since the argument of logarithm, $Vr^2/\alpha$, approaches zero at some $r$ ($\alpha<0$), mainly due to a cancellation of the LO and N$^2$LO contributions.
This behavior is indeed seen in $DD^*({}^3S_1;I=0)$ in Fig.~\ref{fig:Eeff_DDst}, and 
$DD^*({}^3S_1;I=1)$, $D^*D^*({}^3S_1;I=0)$ and $D^*D^*({}^5S_2;I=1)$ in Fig.~\ref{fig:OtherEeff}, with Eqs.~(\ref{VLO}) and~(\ref{VLODstDst}). 
Meanwhile, when the LO OPE is attractive, $E_{\rm eff}$ does not diverge but converges on $m_\pi$ for $r\gg m_\pi^{-1}$. For this reason, the plateau of $E_{\rm eff}\sim 2m_\pi$ for $D^*D^*({}^1S_0;I=1)$ persists to $r\sim 5.0\, {\rm fm}$, which is relatively large, due to its attractive OPE effects. 
When the LO OPE is absent, the relation $E_{\rm eff}=2m_\pi$ exactly holds at large $r$, which is manifestly reflected in $DD({}^1S_0;I=1)$ channel, with Eq.~(\ref{VDDLO}). Therefore, one can conclude that $DD({}^1S_0;I=1)$ is the cleanest channel to see $V\sim \alpha {\rm e}^{-2m_\pi r}/r^2$ and to check the validity of present analysis.



\section{Conclusions}
\label{sec:Conclusions}

In this paper, we have examined TPE effects on the $D$ - $D^*$ potential for the $S$-wave iso-singlet channel, motivated by the HAL QCD lattice results of the form $V^{DD^*}\sim \alpha\left({\rm e}^{-m_\pi}/r\right)^2$ ($\alpha<0$) in the semi-long range of $1.0\, {\rm fm} \lesssim r\lesssim 2.0\, {\rm fm}$, at the nearly physical point $m_\pi=146.4$ MeV. Our analysis is based on the heavy-meson chiral perturbation theory with heavy-quark spin symmetry. We have adopted the dispersion-relation method to derive the potential, for which only (semi-)long-range parts of the $DD^*$ system are well separated with no ambiguities accompanied by regularization scheme or conditions.

We have found that asymptotic behavior of the attractive $V^{DD^*}\propto -\left({\rm e}^{-m_\pi r}/r\right)^2$ ($r\gg m_\pi^{-1}$) is derived at the N$^2$LO TPEs from isospin-independent triangle diagrams, while the NLO TPEs induce repulsive $V^{DD^*}\propto {\rm e}^{-2m_\pi r}/r^{5/2}$, consistently with Ref.~\cite{Wang:2022jop}. It has turned out that, by tuning the unknown ${\cal O}(p^2)$ coupling constants, the lattice result including the behavior of $V^{DD^*}\sim \alpha\left({\rm e}^{-m_\pi}/r\right)^2$ at $1.0\, {\rm fm} \lesssim r\lesssim 2.0\, {\rm fm}$ is reproduced satisfactorily, without introducing any contact interactions and form factors.

In order to check consistency of our findings, we have also presented predictions of the potentials for other $S$-wave $D^{(*)}D^{(*)}$ channels: $DD^*({}^3S_1;I=1)$, $DD({}^1S_0;I=1)$, $D^*D^*({}^1S_0;I=1)$, $D^*D^*({}^3S_1;I=0)$ and $D^*D^*({}^5S_2;I=1)$, with the same set up. In the other channels, we have found that the TPE tail of $V \sim \alpha\left({\rm e}^{-m_\pi}/r\right)^2$ ($\alpha<0$) universally emerges owing to the N$^2$LO contributions in slightly longer regime: $1.5\, {\rm fm} \lesssim r$. Among them, in $DD({}^1S_0;I=1)$, the clear TPE tail is predicted.

Here, we stressed that the present analysis is devoted to delineating the TPE effects at (semi-)long range in the simplest framework, but not to discussing the possibility of bound state or resonance state. For this reason we have neglected the short-range interactions.

Our findings are expected to provide 
useful information on the origin of the TPE tail of the form $V\sim \alpha \left({\rm e}^{-m_\pi r}/r\right)^2$ to future lattice simulations. 
Concurrently we expect future lattice simulations on the various $D^{(*)}D^{(*)}$ potentials, particularly with a larger lattice size, to shed light on the (semi-)long range nature in more detail.

In what follows, we will discuss related topics. In the present work, we have concentrated on (semi-)long regime in $D^{(*)}D^{(*)}$ system to compare with (future) lattice data, and within a description of pion exchanges the HAL QCD data on $D$ - $D^*$ potential is reproduced satisfactorily up to N$^2$LO calculations. With this treatment, the ${\cal O}(p^2)$ coupling constants are predicted to be of ${\cal O}(10^{-1}\, {\rm MeV}^{-1})$, which are substantially larger than those predicted by the resonance saturation model of ${\cal O}(10^{-4}\, {\rm MeV}^{-1})$~\cite{Wang:2022jop}.
In this reference, the authors incorporated contact interactions as well as form factors and other corrections to describe complicated short-range correlations. For better understanding of the TPE tail, it is worth examining those effects with various $D^{(*)}D^{(*)}$ channels. We leave such detailed investigations for future study. 
This extensive research, together with, e.g., quark-exchange effects and coupled-channel effects, is also inevitable to study molecule states of $D^{(*)}D^{(*)}$ more quantitatively. 

Our analysis indicates that the $V\sim \alpha\left({\rm e}^{-m_\pi}/r\right)^2$ behavior in the semi-long range of $S$-wave $D^{(*)}D^{(*)}$ channel is mainly reproduced by isospin-independent triangle diagrams [e.g., (T$1'$) - (T$3'$) diagrams in Fig.~\ref{fig:DDst_NNLO} for $DD^*$ system], as shown in Refs.~\cite{Kaiser:1997mw,Epelbaum:2003gr,Brambilla:2015rqa,TarrusCastella:2018php} for other systems. The similar diagrams can be obtained in $N$ - $Q\bar{Q}$ system, and the lattice result indeed shows a similar TPE tail in the semi-long range regime~\cite{Lyu:2022imf,Lyu:2024ttm}. This fact is consistent with our present conclusion. On the other hand, for $Q\bar{Q}$ - $Q\bar{Q}$ system there is no room to yield the triangle diagrams since the heavy quarkonia $Q\bar{Q}$ carry no isospin. Hence, to check the absence of TPE tail in the semi-long range by future lattice simulations in $Q\bar{Q}$ - $Q\bar{Q}$ system is also expected to provide useful information on its origin.

\section*{Acknowledgement}
\label{sec:Acknowledgement}

We thank Yan Lyu for kindly providing us with the HAL QCD lattice data. We are also grateful to the organizers and participants of Workshop on Heavy-Quark Exotics at Toba for stimulating discussions. We thank Parikshit Junnarkar, Takumi Doi and Yan Lyu for fruitful discussions on the HAL QCD simulations, and Manato Sakai and Yasuhiro Yamaguchi for giving us useful comments.
The work of D.S. is supported in part by Grants-in-Aidfor Scientific Research No.~23K03377No.~23H05439, and No.25K17386. The work of M.H. is supported in part by Grants-in-Aidfor Scientific Research No.~23H05439 and No.~24K07045.

\appendix

\section{One-loop computations for $DD^*$ system}
\label{sec:OneLoops}

In this appendix, we summarize evaluation of the imaginary parts of one-loop amplitudes depicted in Figs.~\ref{fig:DDst_NLO} and~\ref{fig:DDst_NNLO} based on the dimensional regularization. It should be noted that replacement of $d\to4$ may be safely done since here we are only interested in the imaginary parts which are independent of regularization scheme. Here, the momenta are chosen to be
\begin{eqnarray}
&& p'=p+q\ , \ \ p''=p+q_1\ , \nonumber\\
&& p'''=q_2-p\ , \ \ p''''=q_4+p\ ,
\label{def:p}
\end{eqnarray}
as diagrammatically indicated in Fig.~\ref{fig:DDst_NLO}.

Since the total wave function of a boson system must be symmetric, the $DD^*$ states take the forms of Eq.~(\ref{DDstWF}) depending on the isospin $I$. Thus, an additional minus sign emerges in the $u$-channel amplitudes. We attach $(-)^{I+1}$ to take into account this effect. In the following calculations, we will use $v^\mu = (1,{\bm 0})$ and $q^0_1=q^0_2=q^0_3=q^0_4=0$.

\subsection{${\cal O}(\nu^2)$ football diagram (NLO)}
\label{sec:FootBall}

The ${\cal O}(\nu^2)$ football diagram in Fig.~\ref{fig:DDst_NLO} (F1) is evaluated to be
\begin{eqnarray}
{\cal M}^{DD^*}_{F} &=& \frac{1}{f_\pi^4}{\cal C}_I  ({\bm \epsilon}^*\cdot{\bm \epsilon}) L_{F}\ ,
\end{eqnarray}
with a loop function
\begin{eqnarray}
L_{F} &\equiv& \int\frac{d^dp}{i(2\pi)^d}\frac{p_0^2}{(p^2-m_\pi^2)(p'^2-m_\pi^2)}  \nonumber\\
&=&\int\frac{d^dl_E}{(2\pi)^d}\int_0^1dx\frac{-\frac{1}{4}l_E^2}{(l_E^2+\Delta)^2}  \nonumber\\
&=&-\int_0^1dx\frac{\Delta}{2(4\pi)^2}  \ln \Delta + \cdots\ , \label{LFoot}
\end{eqnarray}
where divergences and trivial finite terms are abbreviated in ``$\cdots$'',  and ${\bm \epsilon}$ and ${\bm \epsilon}^*$ are polarization vector of initial- and final-state $D^*$ mesons, respectively. The coefficient ${\cal C}_I$ represents the eigenvalue of an operator $\tau^a\otimes \tau^a$.
The function $\Delta$ is defined by $\Delta= m_\pi^2-x(1-x)s$ with $s=q^2 = -{\bm q}^2$. In Eq.~(\ref{LFoot}) the Feynman-parameter representation is adopted to collect the denominators:
\begin{eqnarray}
\frac{1}{A^\alpha B^\beta} &=& \frac{\Gamma[\alpha+\beta]}{\Gamma[\alpha]\Gamma[\beta]}\int_0^1 dx\frac{x^{\alpha-1}(1-x)^{\beta-1}}{[xA+(1-x)B]^{\alpha+\beta}}\ . \nonumber\\
\end{eqnarray}
The imaginary part of $L_{F}$ is obtained through the replacement of $\Delta\to \Delta-i0$  for which $\ln\Delta \to \ln|\Delta|{\rm e}^{-i\pi}$ when $\Delta<0$. That is
\begin{eqnarray}
{\rm Im} [L_{F} ]&=& \frac{\pi}{2(4\pi)^2}\int_{x_-}^{x_+}dx\, \Delta\nonumber\\
&=& -\frac{\pi}{12(4\pi)^2}\sqrt{1-\frac{4m_\pi^2}{s}} (s-4m_\pi^2) \ ,
\end{eqnarray}
with $x_\pm = \frac{1}{2}\pm\frac{1}{2}\sqrt{1-\frac{4m_\pi^2}{s}}$. We note that this imaginary part appears only above the two-pion threshold: $s>4m_\pi^2$. Thus we can arrive at
\begin{eqnarray}
{\rm Im}[{\cal M}^{DD^*}_{F}] =- \frac{({\bm \epsilon}^*\cdot{\bm \epsilon}){\cal C}_I}{192\pi f_\pi^4}\sqrt{1-\frac{4m_\pi^2}{s}}
(s-4m_\pi^2)\ . \label{ImDDstF}
\end{eqnarray}

\subsection{${\cal O}(\nu^2)$ triangle diagrams (NLO)}
\label{sec:Triangle}

Similarly, the ${\cal O}(\nu^2)$ triangle diagrams in Fig.~\ref{fig:DDst_NLO} (T1) - (T3) are evaluated to be
\begin{eqnarray}
{\cal M}^{DD^*}_{T1} =  \frac{2g^2}{f_\pi^4}{\cal C}_I({\bm \epsilon}^*\cdot{\bm \epsilon})L^{ii}_{T}\ ,
\end{eqnarray}
\begin{eqnarray}
{\cal M}^{DD^*}_{T2} = -\frac{2g^2}{f_\pi^4} {\cal C}_I\epsilon^{*j}\epsilon^i \tilde{L}^{ij}_{T}\ ,
\end{eqnarray}
and
\begin{eqnarray}
{\cal M}^{DD^*}_{T3}= \frac{2g^2}{f_\pi^4}{\cal C}_I\epsilon^{*j}\epsilon^i \left( L^{ii}_{T} \delta^{ij} + \tilde{L}^{ij}_{T} \right)\ ,
\end{eqnarray}
where the loop functions are ($l=p+xq$)
\begin{eqnarray}
L^{ij}_{T} &\equiv& \int\frac{d^dp}{i(2\pi)^d}\frac{p_0p^ip'^j}{(p^2-m_\pi^2)(p'^2-m_\pi^2)(2v\cdot p'')} \ , \nonumber\\
\tilde{L}^{ij}_{T} &\equiv& \int\frac{d^dp}{i(2\pi)^d}\frac{p_0 p^i p'^j}{(p^2-m_\pi^2)(p'^2-m_\pi^2)(2v\cdot p''')}\ . \nonumber\\
\end{eqnarray}
These functions satisfy $\tilde{L}_{T}^{ij} = -L_{T}^{ij}$. Summing up the three amplitudes results in
\begin{eqnarray}
{\cal M}^{DD^*}_T &\equiv& {\cal M}^{DD^*}_{T1} +  {\cal M}^{DD^*}_{T2} +  {\cal M}^{DD^*}_{T3} \nonumber\\
&=&  \frac{4g^2}{f_\pi^4}{\cal C}_I({\bm \epsilon}^*\cdot{\bm \epsilon})L^{ii}_{T}\ .
\end{eqnarray}
The imaginary part of $L^{ij}_{T}$ is calculated to be
\begin{eqnarray}
L^{ij}_{T} &=& \frac{1}{2}\int_0^1dx\int\frac{d^dl_E}{(2\pi)^d}\frac{\frac{1}{4}l_E^2\delta^{ij}-x(1-x)q^iq^j}{(l_E^2+\Delta)^2} \nonumber\\
 &=&  \int_0^1 dx\frac{1}{2(4\pi)^2}\left[\frac{\Delta\delta^{ij}}{2}+x(1-x)q^iq^j\right]{\rm ln}\Delta + \cdots \nonumber\\
&\overset{\rm Im}{\to}&\frac{i}{384\pi}\sqrt{1-\frac{4m_\pi^2}{s}} \nonumber\\
&& \times\Bigg[(s-4m_\pi^2)\delta^{ij} -2(s+2m_\pi^2)\frac{q^iq^j}{s}\Bigg]\ ,
\end{eqnarray}
and thus one can arrive at
\begin{eqnarray}
{\rm Im}[{\cal M}^{DD^*}_{T}] =  {\cal C}_I\frac{g^2}{96\pi f_\pi^4}\sqrt{1-\frac{4m_\pi^2}{s}} (5s-8m_\pi^2)({\bm \epsilon}^*\cdot{\bm \epsilon})\ . \nonumber\\ \label{ImDDstT}
\end{eqnarray}

\subsection{${\cal O}(\nu^2)$ box diagrams (NLO)}
\label{sec:Box}

As for the ${\cal O}(\nu^2)$ box diagrams, first, Fig.~\ref{fig:DDst_NLO} (B1) - (B4) can be evaluated to be
\begin{eqnarray}
{\cal M}^{DD^*}_{B1} = \frac{g^4}{f_\pi^4}{\cal D}_1 \epsilon^{*j}\epsilon^i  L^{ikjk}_{B}\ ,
\end{eqnarray}
\begin{eqnarray}
{\cal M}^{DD^*}_{B2} =  \frac{g^4}{f_\pi^4}{\cal D}_{2} \epsilon^{*j}\epsilon^i  \tilde{L}^{ikjk}_{B}\ ,
\end{eqnarray}
\begin{eqnarray}
{\cal M}^{DD^*}_{B3} = \frac{g^4}{f_\pi^4}{\cal D}_1 \epsilon^{*j}\epsilon^i(L_{B}^{lklk}\delta^{ij}-L^{ikjk}_{B})\ ,
\end{eqnarray}
and
\begin{eqnarray}
{\cal M}^{DD^*}_{B4} = \frac{g^4}{f_\pi^4}{\cal D}_2\epsilon^{*j}\epsilon^i (\tilde{L}_{B}^{lklk}\delta^{ij}-\tilde{L}_{B}^{ikjk})\ ,
\end{eqnarray}
where the loop functions read
\begin{eqnarray}
L^{ijkl}_{B} &\equiv& \int\frac{d^dp}{i(2\pi)^d}\frac{p^ip^j p'^kp'^l}{(p^2-m_\pi^2)(p'^2-m_\pi^2)(2v\cdot p'')(2v\cdot p''')} \ ,  \nonumber\\
\tilde{L}^{ijkl}_{B} &\equiv&\int\frac{d^dp}{i(2\pi)^d}\frac{p^ip^jp'^k p'^l}{(p^2-m_\pi^2)(p'^2-m_\pi^2)(2v\cdot p'')(2v\cdot p'''')}\ , \nonumber\\
\end{eqnarray}
and we have defined the following isospin matrix elements:
\begin{eqnarray}
{\cal D}_1 &\equiv& (\tau^b\tau^a) \otimes(\tau^b\tau^a) \ , \nonumber\\
{\cal D}_2 &\equiv&(\tau^b\tau^a)\otimes (\tau^a\tau^b) \ .
\end{eqnarray}
Now, obviously $\tilde{L}^{ijkl}_{B} = -L^{ijkl}_{B}$ (when Weinberg's pinch singularity~\cite{Weinberg:1990rz,Weinberg:1991um} is appropriately subtracted) for which 
\begin{eqnarray}
{\cal M}^{DD^*}_{B(a)} &\equiv& {\cal M}^{DD^*}_{B1} + {\cal M}^{DD^*}_{B2} + {\cal M}^{DD^*}_{B3} + {\cal M}^{DD^*}_{B4} \nonumber\\
&=& \frac{g^4}{f_\pi^4}({\cal D}_1-{\cal D}_2)L_{B}^{lklk}({\bm \epsilon}^*\cdot{\bm \epsilon})\ . \label{BoxA}
\end{eqnarray}
Utilizing an identity
\begin{eqnarray}
\frac{1}{A^rB^s} = 2^s\frac{\Gamma[r+s]}{\Gamma[r]\Gamma[s]}\int_0^\infty d\lambda\frac{\lambda^{s-1}}{(A+2\lambda B)^{r+s}} \label{CombineHQET}
\end{eqnarray}
to collectively treat the denominators, the loop function $L_{B}^{ijkl}$ is rewritten into
\begin{eqnarray}
L_{B}^{ijkl} &=& -6\int_0^1\int_0^\infty d\lambda\int\frac{d^dl_E}{(2\pi)^d}\frac{\lambda}{(l_E^2+\Delta')^4} \times \nonumber\\
&& \Bigg[\frac{l_E^4}{24}(\delta^{ij}\delta^{kl} + \delta^{ik}\delta^{jl} + \delta^{il}\delta^{jk}) + \frac{1}{4}\Big((1-x)^2\delta^{ij}q^kq^l \nonumber\\
&&-x(1-x)(\delta^{ik}q^jq^l +\delta^{il}q^jq^k +\delta^{jk}q^iq^l +\delta^{jl}q^iq^k ) \nonumber\\
&& + x^2\delta^{kl}q^iq^j\Big) l_E^2 + x^2(1-x)^2q^iq^jq^kq^l \Bigg]\ ,
\end{eqnarray}
where the momentum shift is $l=p+xq+\lambda v$ and $\Delta'=\Delta+\lambda^2$. It should be emphasized that the use of Eq.~(\ref{CombineHQET}) is justified when replacing
\begin{eqnarray}
\frac{1}{(2v\cdot p''+i0)(2v\cdot p'''+i0)} &=& \frac{1}{(2p_0+i0)(-2p_0+i0)} \nonumber\\
&\to& -\frac{1}{(2p_0+i0)^2} \ .
\end{eqnarray}
This procedure is equivalent to the subtraction of 2PR diagrams done in Ref.~\cite{Xu:2017tsr,Xu:2025xrl}. After integrating over the Euclideanized momentum $l_E$, we get
\begin{eqnarray}
L_{B}^{ijkl} &=& \frac{1}{64\pi^2}\int_0^1dx\int_0^\infty d\lambda\, \lambda\ln\Delta'(\delta^{ij}\delta^{kl} + \delta^{ik}\delta^{jl} \nonumber\\
&+&  \delta^{il}\delta^{jk}) -\frac{1}{32\pi^2}\int_0^1dx\int_0^\infty d\lambda\frac{\lambda}{\Delta'}\Bigg[(1-x)^2\delta^{ij}q^kq^l  \nonumber\\
&-&  x(1-x)\Big(\delta^{ik}q^jq^l + \delta^{il}q^jq^k + \delta^{jk}q^iq^l + \delta^{jl}q^iq^k \Big) \nonumber\\
&+&  x^2\delta^{kl}q^iq^j\Bigg]  + \cdots \ . \label{LB1}
\end{eqnarray}
Then, using 
\begin{eqnarray}
\frac{1}{\Delta'-i0} &=& {\rm P}\frac{1}{\Delta'} + i\pi \delta(\Delta') \nonumber\\
&=& {\rm P}\frac{1}{\Delta'}  +  \frac{1}{2\lambda}\delta(\lambda-\lambda_0)
\end{eqnarray}
with $\lambda_0 = \sqrt{-\Delta}$ allows us to evaluate its imaginary part
\begin{eqnarray}
{\rm Im}[L_{B}^{ijkl}] &=&-\frac{i}{768\pi}\Bigg[(s-4m_\pi^2)(\delta^{ij}\delta^{kl} + \delta^{ik}\delta^{jl} +  \delta^{il}\delta^{jk}) \nonumber\\
&+& \frac{ 4(s-m_\pi^2)}{s}(\delta^{ij}q^kq^l + \delta^{kl}q^iq^j) - \frac{2(s+2m_\pi^2)}{s} \nonumber\\
&&  (\delta^{ik}q^jq^l + \delta^{il}q^jq^k + \delta^{jk}q^iq^l + \delta^{jl}q^iq^k)  \Bigg]\nonumber\\
&& \times \sqrt{1-\frac{4m_\pi^2}{s}} \ . \label{ImLB1}
\end{eqnarray}
It should be noted that in deriving Eq.~(\ref{LB1}), one can also find $1/\Delta'^2$ term, but this term does not generate imaginary parts for any $x$ and $\lambda$. Therefore, using
\begin{eqnarray}
{\cal D}_1- {\cal D}_2 = -4{\cal C}_I\ , \label{D1minusD2}
\end{eqnarray}
we can obtain
\begin{eqnarray}
{\rm Im}[{\cal M}^{DD^*}_{B(a)}] = \frac{{\cal C}_Ig^4}{192\pi f_\pi^4}\sqrt{1-\frac{4m_\pi^2}{s}}(23s-20m_\pi^2)({\bm \epsilon}^*\cdot{\bm \epsilon})\ , \nonumber\\ \label{ImDDstBa}
\end{eqnarray}
from Eq.~(\ref{BoxA}).

Next, the remaining box diagrams (B5) and (B6) in Fig.~\ref{fig:DDst_NLO}, which are $u$-channel, are given by
\begin{eqnarray}
{\cal M}^{DD^*}_{B5} =(-)^I \frac{g^4}{f_\pi^4}{\cal D}_1\epsilon^{ikl}\epsilon^{jmn}  \epsilon^{*j}\epsilon^i q^lq^nL^{km}_{B}\ ,
\end{eqnarray}
and
\begin{eqnarray}
{\cal M}^{DD^*}_{B6} = (-)^{I+1}\frac{g^4}{f_\pi^4}{\cal D}_2\epsilon^{ikl}\epsilon^{jmn}  \epsilon^{*j}\epsilon^i q^lq^n\tilde{L}^{km}_{B}\ ,
\end{eqnarray}
where the loop functions are defined by
\begin{eqnarray}
 L_{B}^{km} &=& \int\frac{d^dp}{i(2\pi)^d}\frac{p^k p^m  }{(p^2-m_\pi^2)(p'^2-m_\pi^2)(2v\cdot p'')(2v\cdot p''')} \ ,\nonumber\\
 \tilde{L}_{B}^{km} &=& \int\frac{d^dp}{i(2\pi)^d}\frac{p^k p^m  }{(p^2-m_\pi^2)(p'^2-m_\pi^2)(2v\cdot p'')(2v\cdot p'''')} \ .\nonumber\\
\end{eqnarray}
Here obviously $ L_{B}^{km}  =  -\tilde{L}_{B}^{km} $, and thus
\begin{eqnarray}
{\cal M}^{DD^*}_{B(b)} &\equiv& {\cal M}^{DD^*}_{B5} + {\cal M}^{DD^*}_{B6} \nonumber\\
&=&  (-)^I \frac{g^4}{f_\pi^4}({\cal D}_1+{\cal D}_2)\epsilon^{ikl}\epsilon^{jmn}  \epsilon^{*j}\epsilon^i q^lq^n L^{km}_{B}\ . \nonumber\\ \label{MBoxB}
\end{eqnarray}
 Here, the imaginary part of $L_{B}^{km}$ is easily obtained as
\begin{eqnarray}
L_{B}^{km} 
&=& -\frac{\delta^{km}}{2(4\pi)^2}\int_{x_-}^{x_+}dx\int_0^\infty\frac{\lambda}{\Delta'} + \cdots \nonumber\\
&\overset{\rm Im}{\to}&  -\frac{i\delta^{km}}{64\pi}\sqrt{1-\frac{4m_\pi^2}{s}}\ ,
\end{eqnarray}
which immediately results in
\begin{eqnarray}
{\rm Im}[{\cal M}^{DD^*}_{B(b)}] =   \frac{g^4}{32\pi f_\pi^4}\sqrt{1-\frac{4m_\pi^2}{s}}\Big[ 2s({\bm \epsilon}^*\cdot{\bm \epsilon}) + S_{12}({\bm q})\Big] \ , \nonumber\\ \label{ImDDstBb}
\end{eqnarray}
from Eq.~(\ref{MBoxB}), with $S_{12}({\bm q}) = 3({\bm \epsilon}^*\cdot {\bm q})({\bm \epsilon}\cdot{\bm q})-({\bm \epsilon}^*\cdot{\bm \epsilon}){\bm q}^2$. In obtaining this expression we have made use of
\begin{eqnarray}
{\cal D}_1 + {\cal D}_2 = 6 \ ,
\end{eqnarray}
regardless of the isospin.

In summary, from Eqs.~(\ref{ImDDstF}),~(\ref{ImDDstT}),~(\ref{ImDDstBa}) and~(\ref{ImDDstBb}), the imaginary part at NLO is derived as
\begin{eqnarray}
{\rm Im}[{\cal M}^{DD^*}_{\rm NLO}] &=& \frac{{\cal C}_I}{192\pi f_\pi^4}  \Big[-(s-4m_\pi^2)+2g^2(5s-8m_\pi^2) \nonumber\\
&& +g^4(23s-20m_\pi^2)\Big] \sqrt{1-\frac{4m_\pi^2}{s}} ({\bm \epsilon}^*\cdot{\bm \epsilon})  \nonumber\\
&&  +  \frac{(-)^Ig^4}{32\pi f_\pi^4}\sqrt{1-\frac{4m_\pi^2}{s}}\Big[ 2s({\bm \epsilon}^*\cdot{\bm \epsilon}) + S_{12}({\bm q})\Big] \ . \nonumber\\
\end{eqnarray}

\begin{figure*}[htbp]
\centering
\hspace*{-0.7cm} 
\includegraphics[width=18cm]{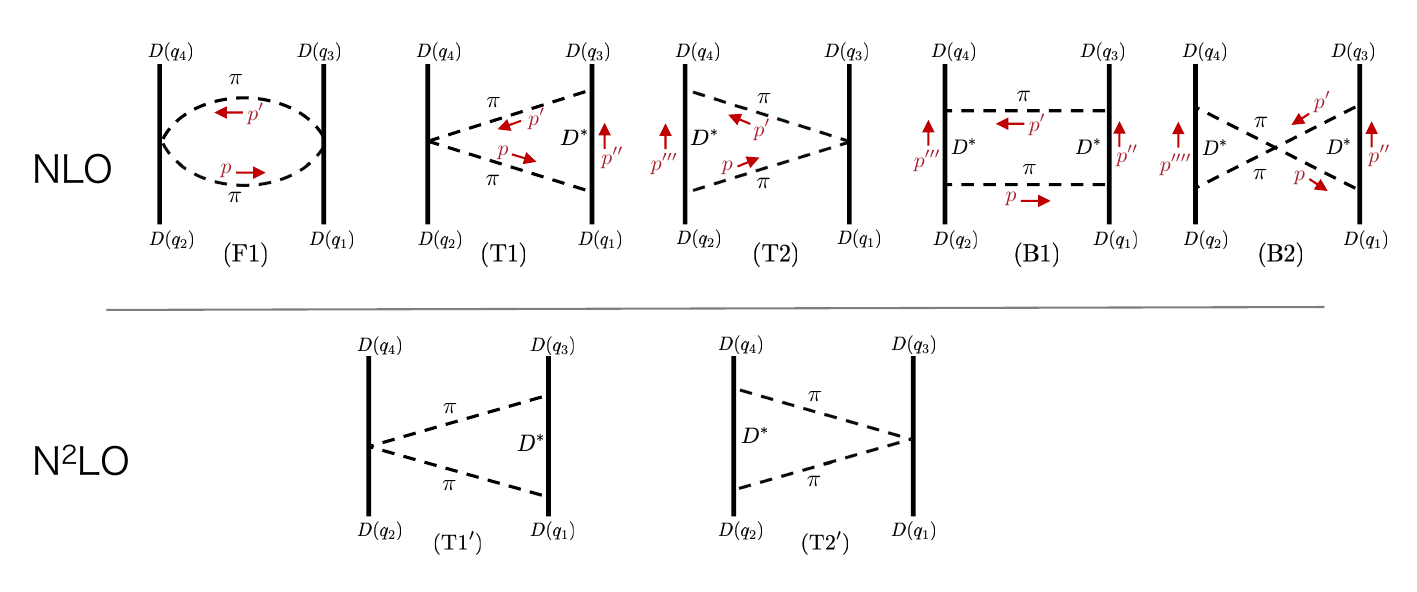}
\caption{NLO and N$^2$LO $DD\to DD$ diagrams generated by TPEs.}
\label{fig:DD_All}
\end{figure*}   

\subsection{${\cal O}(\nu^3)$ triangle diagrams (N$^2$LO)}
\label{sec:NNLOTriangle}

Here we show computations of the ${\cal O}(\nu^3)$ amplitudes. Since those terms generate even numbers of pions, topologically only the triangle diagrams, i.e., (T1$'$) - (T5$'$) in Fig.~\ref{fig:DDst_NNLO} are possible at this order. The corresponding amplitudes read
\begin{eqnarray}
{\cal M}^{DD^*}_{T1'} &=& -\frac{3g^2}{f_\pi^4}({\bm \epsilon}^*\cdot{\bm \epsilon})\Big[h_1L^{(b)ii}_{T'} + h_4 m_\pi^2 L^{(c)ii}_{T'}\Big]  \ , \nonumber\\
\end{eqnarray}
\begin{eqnarray}
{\cal M}^{DD^*}_{T2'} &=& -\frac{3g^2}{f_\pi^4}\epsilon^{*j}\epsilon^i\Big[ h_1\tilde{L}^{(b)ij}_{T'} + h_4 m_\pi^2 \tilde{L}^{(c)ij}_{T'}\Big] \ , \nonumber\\
\end{eqnarray}
\begin{eqnarray}
{\cal M}^{DD^*}_{T3'} &=& -\frac{3g^2}{f_\pi^4}\epsilon^{*j}\epsilon^i\Big[ h_1\left(L^{(b)kk}_{T'}\delta^{ij}-\tilde{L}^{(b)ij}_{T'} \right) \nonumber\\
&&  +h_4 m_\pi^2 \left(L^{(c)kk}_{T'}\delta^{ij}-\tilde{L}^{(c)ij}_{T'} \right)\Big] \ , 
\end{eqnarray}
\begin{eqnarray}
{\cal M}^{DD^*}_{T4'} =(-)^{I+1} \frac{2h_5g^2}{f_\pi^4}{\cal C}_I\epsilon^{*j}\epsilon^i\epsilon^{ikl}\epsilon^{jmn}q^lq^n L_{T'}^{(a)km}\ , \nonumber\\
\end{eqnarray}
and
\begin{eqnarray}
{\cal M}^{DD^*}_{T5'} = (-)^{I+1}\frac{2h_5g^2}{f_\pi^4}{\cal C}_I\epsilon^{*j}\epsilon^i\epsilon^{ikl}\epsilon^{jmn}q^lq^n \tilde{L}_{T'}^{(a)km}\ , \nonumber\\
\end{eqnarray}
where we have omitted trivially vanishing loops. The loop functions are given by
\begin{eqnarray}
L_{T'}^{(a)ij} &\equiv& \int\frac{d^4p}{i(2\pi)^4}\frac{p^ip^j}{(p^2-m_\pi^2)(p'^2-m_\pi^2)(2v\cdot p'')}\ , \nonumber\\ 
L^{(b)ij}_{T'} &\equiv& \int\frac{d^dp}{i(2\pi)^d}\frac{p^i p'^j({\bm p}\cdot{\bm p}')}{(p^2-m_\pi^2)(p'^2-m_\pi^2)(2v\cdot p'')} \, , \nonumber\\
L^{(c)ij}_{T'} &\equiv& \int\frac{d^dp}{i(2\pi)^d}\frac{p^i p'^j}{(p^2-m_\pi^2)(p'^2-m_\pi^2)(2v\cdot p'')} \, , \nonumber\\
\end{eqnarray}
and
\begin{eqnarray}
\tilde{L}_{T'}^{(a)ij} &\equiv& \int\frac{d^4p}{i(2\pi)^4}\frac{p^ip^j}{(p^2-m_\pi^2)(p'^2-m_\pi^2)(2v\cdot p''')} \ , \nonumber\\
\tilde{L}^{(b)ij}_{T'} &\equiv& \int\frac{d^dp}{i(2\pi)^d}\frac{p^i p'^j({\bm p}\cdot{\bm p}')}{(p^2-m_\pi^2)(p'^2-m_\pi^2)(2v\cdot p''')} \, , \nonumber\\
\tilde{L}^{(c)ij}_{T'} &\equiv& \int\frac{d^dp}{i(2\pi)^d}\frac{p^i p'^j}{(p^2-m_\pi^2)(p'^2-m_\pi^2)(2v\cdot p''')} \, , \nonumber\\
\end{eqnarray}
It should be noted that $h_2$ and $h_3$ contributions always vanish since $q_0=0$. Here, one can immediately see $\tilde{L}_{T'}^{(a)ij}  = L_{T'}^{(a)ij} $, $\tilde{L}_{T'}^{(b)ij}  = L_{T'}^{(b)ij} $ and $\tilde{L}_{T'}^{(c)ij}  = L_{T'}^{(c)ij} $. Thus, summing up the three amplitudes leads to
\begin{eqnarray}
{\cal M}^{DD^*}_{\rm N^2LO} & \equiv&{\cal M}^{DD^*}_{T1'} +{\cal M}^{DD^*}_{T2'} + {\cal M}^{DD^*}_{T3'} +{\cal M}^{DD^*}_{T4'} + {\cal M}^{DD^*}_{T5'}\nonumber\\
&=&  -\frac{6g^2}{f_\pi^4}({\bm \epsilon}^*\cdot{\bm \epsilon}) \Big( h_1  L_{T'}^{(b)ii} + h_4m_\pi^2 L_{T'}^{(c)ii} \Big) \nonumber\\
&& + (-)^{I+1} \frac{4h_5g^2}{f_\pi^4}{\cal C}_I\epsilon^{*j}\epsilon^i\epsilon^{ikl}\epsilon^{jmn}q^lq^n L_{T'}^{(a)km} \ . \nonumber\\
\end{eqnarray}
The loop functions $ L_{T'}^{(a)ij} $, $L_{T'}^{(b)ij}$ and $L_{T'}^{(c)ij}$ are reduced to
\begin{eqnarray}
 L_{T'}^{(a)ij}  &=&\frac{2}{(4\pi)^2}\int_0^1dx\int_0^\infty d\lambda\left[\frac{\delta^{ij}}{4}\ln\Delta' - \frac{x^2q^iq^j}{2\Delta'}\right] + \cdots \nonumber\\
&\overset{\rm Im}{\to}& -\frac{i}{256}\frac{1}{\sqrt{s}}\left[(s-4m_\pi^2)\delta^{ij} + (3s-4m_\pi^2)\frac{q^iq^j}{s}\right]\ ,  \label{LT1Primea} 
\nonumber\\
\end{eqnarray}
\begin{eqnarray}
L_{T'}^{(b)ij}  
&=&- \frac{2}{(4\pi)^2}\int_0^1 dx\int_0^\infty d\lambda\Big[\frac{5\Delta'\delta^{ij}}{8}\ln\Delta'  \nonumber\\
&& +  \frac{(x^2-x)s\delta^{ij} - (7x^2-7x+1)q^iq^j}{4}\ln\Delta'  \nonumber\\
&&  -\frac{x^2(1-x)^2sq^iq^j}{2\Delta'} \Big]  + \cdots\nonumber\\
&\overset{\rm Im}{\to}&- \frac{i}{512}\frac{(s-2m_\pi^2)}{\sqrt{s}} \nonumber\\
&& \times \left[ (s-4m_\pi^2)\delta^{ij}-(s+4m_\pi^2)\frac{q^iq^j}{s}\right] \ , \label{LT1Primeb}
\end{eqnarray}
and
\begin{eqnarray}
L_{T'}^{(c)ij} &=&-\frac{2}{(4\pi)^2}\int_0^1 dx\int_0^\infty d\lambda\Big[-\frac{\delta^{ij}}{4}\ln\Delta'  \nonumber\\
&& - \frac{x(1-x)q^iq^j}{2\Delta'} \Big]  + \cdots \nonumber\\
&\overset{\rm Im}{\to}& - \frac{i}{256}\frac{1}{\sqrt{s}}\left[ (s-4m_\pi^2)\delta^{ij}-(s+4m_\pi^2)\frac{q^iq^j}{s}\right]\ , \nonumber\\
\label{LT1Primec}
\end{eqnarray}
respectively. Therefore, finally we arrive at
\begin{eqnarray}
&& {\rm Im}[{\cal M}^{DD^*}_{\rm N^2LO}] \nonumber\\
 &=& \frac{3g^2}{64f_\pi^4} \frac{h_1(s-2m_\pi^2)^2 + 2h_4m_\pi^2(s-2m_\pi^2)}{\sqrt{s}}({\bm \epsilon}^*\cdot{\bm \epsilon}) \nonumber\\
&& + (-)^{I+1} \frac{{\cal C}_Ih_5g^2}{192 f_\pi^4}\frac{s-4m_\pi^2}{\sqrt{s}} \Big[2s ({\bm \epsilon}^*\cdot{\bm \epsilon}) +S_{12}({\bm q})\Big] \ . \nonumber\\
\end{eqnarray}

\begin{figure*}[htbp]
\centering
\hspace*{-0.7cm} 
\includegraphics[width=18cm]{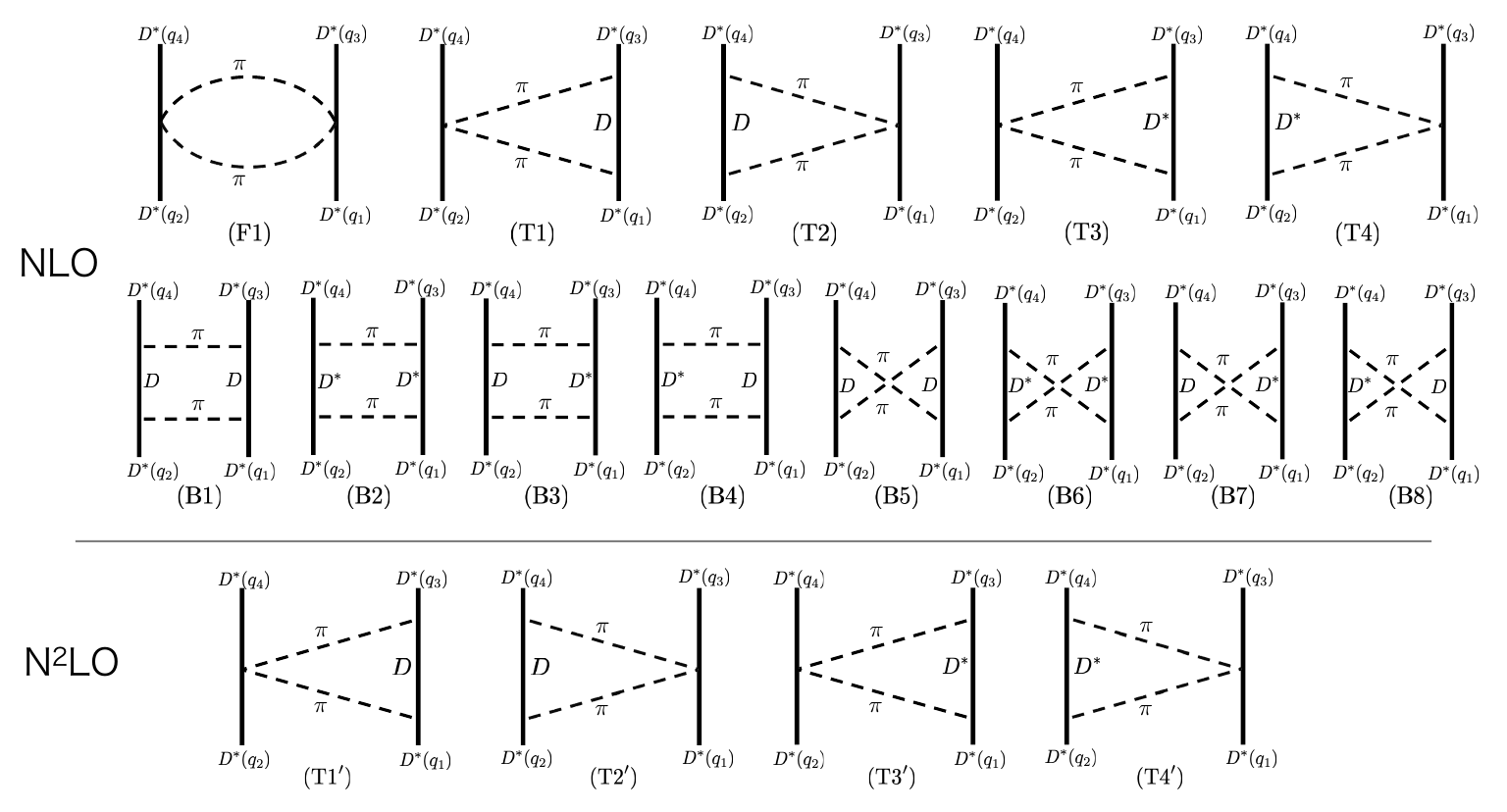}
\caption{NLO and N$^2$LO $D^*D^*\to D^*D^*$ diagrams generated by TPEs.}
\label{fig:DstDst_All}
\end{figure*}   

\section{Computations of $DD$ potential}
\label{sec:DDOneLoops}

Here we present the potential for $S$-wave $DD$ system: $DD({}^1S_0;I=1)$. In this channel, the OPE does not contribute since there is no $DD\pi$ coupling, so that only NLO and N$^2$LO TPEs exist. The corresponding TPE diagrams are summarized in Fig.~\ref{fig:DD_All}.

First, we show results from the NLO TPEs. The corresponding football, triangle and box diagrams are calculated as
\begin{eqnarray}
{\cal M}^{DD}_{F} = \frac{1}{f_\pi^4}{\cal C}_I L_{F}\ ,
\end{eqnarray}
\begin{eqnarray}
{\cal M}^{DD}_{T1}  &=&  \frac{2g^2}{f_\pi^4}{\cal C}_IL^{ii}_{T} \ , \nonumber\\
{\cal M}^{DD}_{T2}  &=& - \frac{2g^2}{f_\pi^4}{\cal C}_I\tilde{L}^{ii}_{T}  \ ,
\end{eqnarray}
and
\begin{eqnarray}
{\cal M}^{DD}_{B1} &=& \frac{g^4}{f_\pi^4}{\cal D}_{1}L^{ikik}_{B} \ , \nonumber\\
{\cal M}^{DD}_{B2} &=& \frac{g^4}{f_\pi^4}{\cal D}_{2} \tilde{L}^{ikik}_{B}  \ ,
\end{eqnarray}
respectively, where the isospin factor is always ${\cal C}_I=+1$ since only $I=1$ is allowed. In these equations, all the loop functions are already computed in Appendix~\ref{sec:OneLoops}, and thus utilizing the previous results immediately yields
\begin{eqnarray}
{\rm Im}[{\cal M}^{DD}_{F}] =- \frac{{\cal C}_I}{192\pi f_\pi^4}\sqrt{1-\frac{4m_\pi^2}{s}}(s-4m_\pi^2)\ ,
\end{eqnarray}
\begin{eqnarray}
{\rm Im}[{\cal M}_T^{DD}] &=& {\rm Im}[{\cal M}^{DD}_{T1} ] + {\rm Im}[{\cal M}^{DD}_{T2} ] \nonumber\\
&=&  \frac{{\cal C}_Ig^2}{96\pi f_\pi^4}(5s-8m_\pi^2)\sqrt{1-\frac{4m_\pi^2}{s}}\  ,
\end{eqnarray}
and
\begin{eqnarray}
{\rm Im}[M_{B}^{DD}] &=& {\rm Im}[M_{B1}^{DD}] + {\rm Im}[M_{B2}^{DD}] \nonumber\\
&=&  \frac{{\cal C}_Ig^4}{192\pi f_\pi^4}(23s-20m_\pi^2)\sqrt{1-\frac{4m_\pi^2}{s}} \ ,\ 
\end{eqnarray}
with the relation~(\ref{D1minusD2}). Thus, summing up the three parts yields 
\begin{eqnarray}
{\rm Im}[{\cal M}^{DD}_{\rm NLO}]  &=& \frac{{\cal C}_I}{192\pi f_\pi^4}\Big[-(s-4m_\pi^2) + 2g^2(5s-8m_\pi^2) \nonumber\\
&& + g^4(23s-20m_\pi^2) \Big]\sqrt{1-\frac{4m_\pi^2}{s}} \ . \label{ImDDNLO}
\end{eqnarray}

Next, we proceed with evaluation of the N$^2$LO TPE diagrams depicted in Fig.~\ref{fig:DD_All}. The corresponding amplitudes read
\begin{eqnarray}
{\cal M}_{T1'} &=& - \frac{3g^2h_1}{f_\pi^4} L_{T'}^{(b)ii} - \frac{3g^2h_4m_\pi^2}{f_\pi^4}  L_{T'}^{(c)ii}\ , \nonumber\\
{\cal M}_{T2'} &=& - \frac{3g^2h_1}{f_\pi^4} \tilde{L}_{T'}^{(b)ii} - \frac{3g^2h_4m_\pi^2}{f_\pi^4}  \tilde{L}_{T'}^{(c)ii}\ ,
\end{eqnarray}
where we have omitted trivially vanishing contributions. Therefore, using the results in Eqs.~(\ref{LT1Primeb}) and~(\ref{LT1Primec}), it follows that
\begin{eqnarray}
{\rm Im}[{\cal M}^{DD}_{\rm N^2LO}] &=&  {\rm Im}[{\cal M}^{DD}_{T1'}]  + {\rm Im}[{\cal M}^{DD}_{T2'}]  \nonumber\\
&=& \frac{3g^2}{64f_\pi^4}\frac{h_1(s-2m_\pi^2)^2 + 2h_4m_\pi^2(s-2m_\pi^2)}{\sqrt{s}}\ . \nonumber\\ \label{ImDDNNLO}
\end{eqnarray}

From Eqs.~(\ref{ImDDNLO}) and~(\ref{ImDDNNLO}), the $DD$ potential is derived to be
\begin{eqnarray}
&& V^{DD}_{\rm NLO}  \nonumber\\
&=& \frac{{\cal C}_Im_\pi}{128\pi^3 f_\pi^4 r^4}\Big[(1-10g^2-23g^4)m_\pi r K_0(2m_\pi r) + \nonumber\\
&&   \Big(1-10g^2-23g^4  - 4g^2(1+3g^2)m_\pi^2 r^2\Big)K_1(2m_\pi r)\Big] \ , \nonumber\\
\end{eqnarray}
and
\begin{eqnarray}
&&V^{DD}_{\rm N^2LO} \nonumber\\
&=&  -\frac{3g^2}{128\pi^2 f_\pi^4 r^6}{\rm e}^{-2m_\pi r}\Big[  h_1 (6+12m_\pi r+10m_\pi^2r^2\nonumber\\
&&  +4m_\pi^3 r^3 + m_\pi^4 r^4) + h_4m_\pi^2 r^2(1+m_\pi r)^2 \Big] \ , 
\end{eqnarray}
while $V_{\rm LO}^{DD}=0$, with the help of the dispersion-relation formula~(\ref{Potential}).

\section{Computations of $D^*D^*$ potential}
\label{sec:DstDstOneLoops}

Here we present the potential for $S$-wave $D^*D^*$ system. In this channel, three types of states are possible:
\begin{eqnarray}
 && D^*D^*({}^1S_0; I=1)\ , \nonumber\\
 &&  D^*D^*({}^3S_1; I=0)\ , \nonumber\\
&& D^*D^*({}^5S_2; I=1)\ .
\end{eqnarray}
As for the polarization vector of $D^*$ mesons, we may employ the following combinations with, e.g., total spin $J$ and its $z$-component $j_z$:
\\
\begin{itemize}
\item \underline{For $D^*D^*({}^1S_0; I=1)$}
\end{itemize}
\begin{eqnarray}
 \frac{1}{\sqrt{3}} {\bm \epsilon}_{+1} \otimes{\bm \epsilon}_{-1} - \frac{1}{\sqrt{3}} {\bm \epsilon}_{0} \otimes{\bm \epsilon}_{0} +  \frac{1}{\sqrt{3}} {\bm \epsilon}_{-1} \otimes{\bm \epsilon}_{+1}
\end{eqnarray}
with $J=0,j_z=0$,
\begin{itemize}
\item \underline{For $D^*D^*({}^3S_1; I=0)$}
\end{itemize}
\begin{eqnarray}
 \frac{1}{\sqrt{2}} {\bm \epsilon}_{+1} \otimes{\bm \epsilon}_{0} - \frac{1}{\sqrt{2}} {\bm \epsilon}_{0} \otimes{\bm \epsilon}_{+1}
\end{eqnarray}
with $J=1,j_z=+1$,
\begin{itemize}
\item \underline{For $D^*D^*({}^5S_2; I=1)$}
\end{itemize}
\begin{eqnarray}
{\bm \epsilon}_{+1} \otimes{\bm \epsilon}_{+1} 
\end{eqnarray}
with $J=2,j_z=+2$. Within the spherical basis, the polarization vector are given by
\begin{eqnarray}
{\bm \epsilon}_{\pm 1} = \frac{1}{\sqrt{2}}(\mp 1,-i,0)\ , \ \ {\bm \epsilon}_0 =(0,0,1)\ .
\end{eqnarray}

For $D^*D^*$ system, the LO OPE as well as the NLO and N$^2$LO TPEs contribute. The LO OPE amplitude for $D^*D^*$ system is readily calculated to be
\begin{eqnarray}
{\cal M}^{D^*D^*}_{\rm LO} &=& \frac{{\cal C}_Ig^2}{3f_\pi^2}\Big[ \big({\bm \epsilon}_{(1)}\cdot{\bm \epsilon}_{(2)}\big) \big({\bm \epsilon}^*_{(1)}\cdot{\bm \epsilon}^*_{(2)}\big) \nonumber\\
&& -  \big({\bm \epsilon}^*_{(2)}\cdot{\bm \epsilon}_{(1)}\big) \big({\bm \epsilon}^*_{(1)}\cdot{\bm \epsilon}_{(2)}\big) \Big] \frac{m_\pi^2}{{\bm q}^2+m_\pi^2}\ , \label{MDstDstLO}
\end{eqnarray}
where we have attached the subscripts (1) and (2) to simply distinguish two $D^*$ mesons in the initial and final states. The potential is straightforwardly obtained via the Fourier transformation.

Similarly to the computations for $DD^*$ and $DD$ systems, the NLO and N$^2$LO TPE diagrams, which are displayed in Fig.~\ref{fig:DstDst_All}, are readily evaluated. The NLO football, triangle, and box diagrams read
\begin{eqnarray}
{\cal M}^{D^*D^*}_{F} = \frac{{\cal C}_I}{f_\pi^4} \big({\bm \epsilon}^*_{(1)}\cdot{\bm \epsilon}_{(1)}\big) \big({\bm \epsilon}^*_{(2)}\cdot{\bm \epsilon}_{(2)}\big) L_F\ ,
\end{eqnarray}
\begin{eqnarray}
{\cal M}^{D^*D^*}_{T1}  =  \frac{2g^2}{f_\pi^4}{\cal C}_I \big({\bm \epsilon}^*_{(2)}\cdot{\bm \epsilon}_{(2)}\big) \epsilon^{*j}_{(1)}\epsilon_{(1)}^i L_{T}^{ij} \ ,
\end{eqnarray}
\begin{eqnarray}
{\cal M}^{D^*D^*}_{T2} = -\frac{2g^2}{f_\pi^4}{\cal C}_I\big({\bm \epsilon}^*_{(1)}\cdot{\bm \epsilon}_{(1)}\big)  \epsilon^{*j}_{(2)}\epsilon_{(2)}^i \tilde{L}_{T}^{ij}\ ,
\end{eqnarray}
\begin{eqnarray}
{\cal M}^{D^*D^*}_{T3} &=& -\frac{2g^2}{f_\pi^4}{\cal C}_I \big({\bm \epsilon}^*_{(2)}\cdot{\bm \epsilon}_{(2)}\big) \nonumber\\
&& \times  \Big[  \epsilon^{*j}_{(1)}\epsilon_{(1)}^i L_{T}^{ji} - \big({\bm \epsilon}^*_{(1)}\cdot{\bm \epsilon}_{(1)}\big)L^{ii}_{T}\Big] \, ,
\end{eqnarray}
\begin{eqnarray}
{\cal M}^{D^*D^*}_{T4} &=& \frac{2g^2}{f_\pi^4}{\cal C}_I\big({\bm \epsilon}^*_{(1)}\cdot{\bm \epsilon}_{(1)}\big) \nonumber\\
&& \times \Big[ \epsilon^{*j}_{(2)}\epsilon_{(2)}^i \tilde{L}_{T}^{ji} -  \big({\bm \epsilon}^*_{(2)}\cdot{\bm \epsilon}_{(2)}\big) \tilde{L}^{ii}_{T} \Big]  \, ,
\end{eqnarray}
\begin{eqnarray}
{\cal M}^{D^*D^*}_{B1} = \frac{g^4}{f_\pi^4}{\cal D}_1 \epsilon^{*l}_{(2)}\epsilon_{(2)}^j \epsilon^{*k}_{(1)}\epsilon_{(1)}^i L_{B}^{ijkl} \ ,
\end{eqnarray}
\begin{eqnarray}
{\cal M}^{D^*D^*}_{B2} &=& \frac{g^4}{f_\pi^4}{\cal D}_1 \Big[ \epsilon^{*l}_{(2)}\epsilon_{(2)}^j \epsilon^{*k}_{(1)}\epsilon_{(1)}^i  L^{klij}_{B} - \big({\bm \epsilon}^*_{(2)}\cdot{\bm \epsilon}_{(2)}\big) \nonumber\\
&& \times \epsilon^{*j}_{(1)}\epsilon_{(1)}^i L_{B}^{jkik} - \big({\bm \epsilon}^*_{(1)}\cdot{\bm \epsilon}_{(1)}\big)  \epsilon^{*j}_{(2)}\epsilon_{(2)}^i  L_{B}^{jkik} \nonumber\\
&& +\big({\bm \epsilon}^*_{(1)}\cdot{\bm \epsilon}_{(1)}\big) \big({\bm \epsilon}^*_{(2)}\cdot{\bm \epsilon}_{(2)}\big) L^{ikik}_{B} \Big] \ ,
\end{eqnarray}
\begin{eqnarray}
{\cal M}^{D^*D^*}_{B3} &=&- \frac{g^4}{f_\pi^4}{\cal D}_1\Big[ \epsilon^{*l}_{(2)}\epsilon_{(2)}^j \epsilon^{*k}_{(1)}\epsilon_{(1)}^i   L_{B}^{jkil} \nonumber\\
&& - \big({\bm \epsilon}^*_{(1)}\cdot{\bm \epsilon}_{(1)}\big)  \epsilon^{*j}_{(2)}\epsilon_{(2)}^i  L_{B}^{ikjk} \Big]\ ,
\end{eqnarray}
\begin{eqnarray}
{\cal M}^{D^*D^*}_{B4} &=&- \frac{g^4}{f_\pi^4}{\cal D}_1\Big[ \epsilon^{*l}_{(2)}\epsilon_{(2)}^j \epsilon^{*k}_{(1)}\epsilon_{(1)}^i  L^{iljk}_{B} \nonumber\\
&& -  \big({\bm \epsilon}^*_{(2)}\cdot{\bm \epsilon}_{(2)}\big)   \epsilon^{*j}_{(1)}\epsilon_{(1)}^i  L_{B}^{ikjk} \Big] \ ,
\end{eqnarray}
\begin{eqnarray}
{\cal M}^{D^*D^*}_{B5} = \frac{g^4}{f_\pi^4}{\cal D}_2 \epsilon^{*l}_{(2)}\epsilon_{(2)}^j \epsilon^{*k}_{(1)}\epsilon_{(1)}^i \tilde{L}_{B}^{ilkj} \ ,
\end{eqnarray}
\begin{eqnarray}
{\cal M}^{D^*D^*}_{B6} &=& \frac{g^4}{f_\pi^4}{\cal D}_2 \Big[ \epsilon^{*l}_{(2)}\epsilon_{(2)}^j \epsilon^{*k}_{(1)}\epsilon_{(1)}^i  \tilde{L}^{kjil}_{B} - \big({\bm \epsilon}^*_{(2)}\cdot{\bm \epsilon}_{(2)}\big) \nonumber\\
&& \times \epsilon^{*j}_{(1)}\epsilon_{(1)}^i \tilde{L}_{B}^{jkik} - \big({\bm \epsilon}^*_{(1)}\cdot{\bm \epsilon}_{(1)}\big)  \epsilon^{*j}_{(2)}\epsilon_{(2)}^i  \tilde{L}_{B}^{ikjk} \nonumber\\
&& + \big({\bm \epsilon}^*_{(1)}\cdot{\bm \epsilon}_{(1)}\big) \big({\bm \epsilon}^*_{(2)}\cdot{\bm \epsilon}_{(2)}\big) \tilde{L}^{ikik}_{B} \Big] \ ,
\end{eqnarray}
\begin{eqnarray}
{\cal M}^{D^*D^*}_{B7} &=&- \frac{g^4}{f_\pi^4}{\cal D}_2\Big[ \epsilon^{*l}_{(2)}\epsilon_{(2)}^j \epsilon^{*k}_{(1)}\epsilon_{(1)}^i   \tilde{L}_{B}^{lkij} \nonumber\\
&& - \big({\bm \epsilon}^*_{(1)}\cdot{\bm \epsilon}_{(1)}\big)  \epsilon^{*j}_{(2)}\epsilon_{(2)}^i  \tilde{L}_{B}^{jkik} \Big]\ ,
\end{eqnarray}
\begin{eqnarray}
{\cal M}^{D^*D^*}_{B8} &=&- \frac{g^4}{f_\pi^4}{\cal D}_2\Big[ \epsilon^{*l}_{(2)}\epsilon_{(2)}^j \epsilon^{*k}_{(1)}\epsilon_{(1)}^i  \tilde{L}^{ijlk}_{B} \nonumber\\
&& -  \big({\bm \epsilon}^*_{(2)}\cdot{\bm \epsilon}_{(2)}\big)   \epsilon^{*j}_{(1)}\epsilon_{(1)}^i  \tilde{L}_{B}^{ikjk} \Big] \ ,
\end{eqnarray}
where all the loop functions are already computed in Appendix~\ref{sec:OneLoops}. Therefore, using the previous results, one can arrive at
\begin{eqnarray}
{\rm Im}[{\cal M}_F^{D^*D^*}] &=& -\frac{{\cal C}_I}{192\pi f_\pi^4} ({\bm \epsilon}^*_{(1)}\cdot{\bm \epsilon}_{(1)}) ({\bm \epsilon}_{(2)}^*\cdot{\bm \epsilon}_{(2)}) \nonumber\\
&& \times (s-4m_\pi^2)\sqrt{1-\frac{4m_\pi^2}{s}}\ ,
\end{eqnarray}
\begin{eqnarray}
{\rm Im}[{\cal M}_T^{D^*D^*}] &=& \frac{{\cal C}_Ig^2}{96\pi f_\pi^4}({\bm \epsilon}^*_{(1)}\cdot{\bm \epsilon}_{(1)}) ({\bm \epsilon}_{(2)}^*\cdot{\bm \epsilon}_{(2)}) \nonumber\\
&& \times (5s-8m_\pi^2)\sqrt{1-\frac{4m_\pi^2}{s}}\ ,
\end{eqnarray}
\begin{eqnarray}
{\rm Im}[{\cal M}_B^{D^*D^*}] &=& \frac{g^4}{16\pi f_\pi^4}\Big[({\bm \epsilon}_{(1)}\cdot{\bm \epsilon}_{(2)})({\bm \epsilon}_{(1)}^*\cdot{\bm \epsilon}_{(2)}^*) \nonumber\\
&& -({\bm \epsilon}_{(1)}^*\cdot{\bm \epsilon}_{(2)})({\bm \epsilon}_{(2)}^*\cdot {\bm \epsilon}_{(1)} )\Big] s\sqrt{1-\frac{4m_\pi^2}{s}} \nonumber\\
&& + \frac{{\cal C}_Ig^4}{192\pi f_\pi^4}({\bm \epsilon}^*_{(1)}\cdot{\bm \epsilon}_{(1)})({\bm \epsilon}^*_{(2)}\cdot{\bm \epsilon}_{(2)}) \nonumber\\
&& \times (23s-20m_\pi^2)\sqrt{1-\frac{4m_\pi^2}{s}} + \cdots\ ,  \nonumber\\
\label{MBDstDst}
\end{eqnarray}
where ${\cal M}_T^{D^*D^*}$ and ${\cal M}_B^{D^*D^*}$ stand for the sums of all triangle and box diagrams, respectively. We note that the dots in Eq.~(\ref{MBDstDst}) represent contributions from the tensor operator which are irrelevant as long as we stick to the $S$-wave $D^*D^*$ state. Thus, summing up the three imaginary parts, 
\begin{eqnarray} 
 {\rm Im}[{\cal M}_{\rm NLO}^{D^*D^*}] &=& \frac{{\cal C}_I}{192\pi f_\pi^4} \Big[ -(s-4m_\pi^2)  + 2g^2(5s-8m_\pi^2) \nonumber\\
 &&+ g^4(23s-20m_\pi^2)\Big]\sqrt{1-\frac{4m_\pi^2}{s}}  \nonumber\\
 && -{\cal C}_J \frac{g^4}{16\pi f_\pi^4}s\sqrt{1-\frac{4m_\pi^2}{s}}\ , \label{ImDstDstNLO}
\end{eqnarray}
where the coefficient ${\cal C}_J$ is
\begin{eqnarray}
{\cal C}_J = \left\{
\begin{array}{cc}
-2 & ({\rm for}\ {}^1S_0) \\
-1 & ({\rm for}\ {}^3S_1) \\
+1 & ({\rm for}\ {}^5S_2) \\
\end{array}
\right. \ .
\end{eqnarray}

Finally, the N$^2$LO triangle diagrams displayed in Fig.~\ref{fig:DstDst_All} are collectively evaluated to be
\begin{eqnarray}
{\cal M}^{D^*D^*}_{\rm N^2LO} &=& {\cal M}^{D^*D^*}_{T1'} + {\cal M}^{D^*D^*}_{T2'} + {\cal M}^{D^*D^*}_{T3'} + {\cal M}^{D^*D^*}_{T4'} \nonumber\\
&=& -  \frac{3g^2}{f_\pi^4}  \big({\bm \epsilon}^*_{(2)}\cdot{\bm \epsilon}_{(2)}\big) \epsilon_{(1)}^{*j}\epsilon_{(1)}^i(h_1L^{(b)ij}_{T'} +h_4m_\pi^2 L^{(c)ij}_{T'}) \nonumber\\
&& - \frac{2{\cal C}_Ig^2h_5}{f_\pi^4} \epsilon^{*l}_{(2)}\epsilon_{(2)}^j \epsilon^{*k}_{(1)}\epsilon_{(1)}^i  \big(L^{ijk}_{T'} q^l-L^{ikl}_{T'} q^j \nonumber\\
&& + L_{T'}^{(a)ij}q^kq^l-L_{T'}^{(a)il}q^jq^k\big)   \nonumber\\
&& -  \frac{3g^2}{f_\pi^4}  \big({\bm \epsilon}^*_{(1)}\cdot{\bm \epsilon}_{(1)}\big) \epsilon_{(2)}^{*j}\epsilon_{(2)}^i (h_1\tilde{L}^{(b)ij}_{T'} + h_4m_\pi^2 \tilde{L}^{(c)ij}_{T'}) \nonumber\\
&& - \frac{2{\cal C}_Ig^2h_5}{f_\pi^4}\epsilon^{*l}_{(2)}\epsilon_{(2)}^j \epsilon^{*k}_{(1)}\epsilon_{(1)}^i  \big(\tilde{L}^{ijl}_{T'} q^k-\tilde{L}^{jkl}_{T'} q^i \nonumber\\
&& + \tilde{L}_{T'}^{(a)ij}q^kq^l-\tilde{L}_{4'}^{(a)jk}q^iq^l\big) \ ,
\end{eqnarray}
where we have newly defined the following loop functions:
\begin{eqnarray}
L^{ijk}_{T'} &\equiv& \int\frac{d^dp}{i(2\pi)^d}\frac{p^i p^j p^k}{(p^2-m_\pi^2)(p'^2-m_\pi^2)(2v\cdot p'')} \ , \nonumber\\
\tilde{L}^{ijk}_{T'} &\equiv& \int\frac{d^dp}{i(2\pi)^d}\frac{p^i p^j p^k}{(p^2-m_\pi^2)(p'^2-m_\pi^2)(2v\cdot p''')} \ , \nonumber\\
\end{eqnarray}
with $\tilde{L}^{ijk}_{T'}  = L^{ijk}_{T'} $. This loop function is reduced to
\begin{eqnarray}
L^{ijk}_{T'} &=& -\frac{1}{(4\pi)^{2}}\int_0^1dx\int_0^\infty d\lambda\Bigg[(xq^k\delta^{ij} + xq^j\delta^{ki} \nonumber\\
&& + xq^i\delta^{jk}) \ln\Delta' - \frac{2x^3q^iq^jq^k}{\Delta'}\Bigg] + \cdots \nonumber\\
&\overset{\rm Im}{\to}&  \frac{i}{256} \Bigg[(q^k\delta^{ij} + q^j\delta^{ki} + q^i\delta^{jk})\frac{s-4m_\pi^2}{\sqrt{s}} \nonumber\\
&& +2q^iq^jq^k\frac{5s-12m_\pi^2}{s^{3/2}}\Bigg] \ ,
\end{eqnarray}
which results in
\begin{eqnarray}
{\rm Im}[{\cal M}^{D^*D^*}_{\rm N^2LO}]  &=& \frac{g^2h_1}{64 f_\pi^4}\frac{(s-2m_\pi^2)^2}{\sqrt{s}}({\bm \epsilon}_{(1)}^*\cdot{\bm \epsilon}_{(1)}) ({\bm \epsilon}_{(2)}^*\cdot{\bm \epsilon}_{(2)})  \nonumber\\
&+&  \frac{g^2h_4}{32 f_\pi^4}\frac{(s-2m_\pi^2)m_\pi^2}{\sqrt{s}} ({\bm \epsilon}_{(1)}^*\cdot{\bm \epsilon}_{(1)}) ({\bm \epsilon}_{(2)}^*\cdot{\bm \epsilon}_{(2)}) \nonumber\\
&+&  \frac{{\cal C}_Ig^2h_5}{384 f_\pi^4}\frac{s(s-4m_\pi^2)}{\sqrt{s}}\Big(({\bm \epsilon}_{(1)}^*\cdot{\bm \epsilon}_{(2)})({\bm \epsilon}_{(2)}^*\cdot{\bm \epsilon}_{(1)}) \nonumber\\
&-&  ({\bm \epsilon}_{(1)}^*\cdot{\bm \epsilon}^*_{(2)})({\bm \epsilon}_{(1)}\cdot{\bm \epsilon}_{(2)}) \Big)  + \cdots\ .
\end{eqnarray}
Namely, the imaginary part reads
\begin{eqnarray}
{\rm Im}[{\cal M}^{D^*D^*}_{\rm N^2LO}] &=& \frac{g^2}{64f_\pi^4}\frac{h_1(s-2m_\pi^2)^2 + 2h_4m_\pi^2(s-2m_\pi^2)}{\sqrt{s}} \nonumber\\
&& + \frac{{\cal C}_I{\cal C}_Jg^2h_5}{384 f_\pi^4}\frac{s(s-4m_\pi^2)}{\sqrt{s}}\ . \label{ImDstDstNNLO}
\end{eqnarray}
Therefore, using the dispersion formula~(\ref{Potential}), the $D^*D^*$ potential for ${}^1S_0$, ${}^3S_1$ and ${}^5S_2$ is derived to be
\begin{eqnarray}
&& V^{D^*D^*}_{\rm NLO}\nonumber\\
&=& \frac{{\cal C}_Im_\pi}{128 \pi^3 f_\pi^4 r^4}\Big[\Big(1 - 10 g^2-23g^4\Big) m_\pi r K_0(2m_\pi r) +\nonumber\\
&& \Big(1- 10g^2 - 23g^4 - 4\big(g^2  + 3g^4\big)m_\pi^2r^2\Big)  K_1(2m_\pi r)\Big]
\nonumber\\
&& + \frac{{\cal C}_Jg^4m_\pi}{32 \pi^3 f_\pi^4 r^4}\Big[3 m_\pi r K_0(2m_\pi r)  \nonumber\\
&& + (3 +2m_\pi^2r^2)  K_1(2m_\pi r)\Big] \ ,
\end{eqnarray}
and
\begin{eqnarray}
&& V^{D^*D^*}_{\rm N^2LO} \nonumber\\
&=&  -\frac{g^2}{128\pi^2 f_\pi^4 r^6}{\rm e}^{-2m_\pi r}\Big[  h_1 (6+12m_\pi r+10m_\pi^2r^2 \nonumber\\
&& +4m_\pi^3 r^3 + m_\pi^4 r^4) + h_4m_\pi^2 r^2(1+m_\pi r)^2 \Big]  \nonumber\\
&& - \frac{{\cal C}_I{\cal C}_Jg^2h_5}{384\pi^2 f_\pi^4 r^6}{\rm e}^{-2m_\pi r}\left(3+6m_\pi r+5m_\pi^2 r^2+2m_\pi^3 r^3\right)  \ , \nonumber\\
\end{eqnarray}
from the corresponding imaginary parts~(\ref{ImDstDstNLO}) and~(\ref{ImDstDstNNLO}), while the LO OPE potential reads
\begin{eqnarray}
V_{\rm LO}^{D^*D^*} &=& \frac{{\cal C}_I{\cal C}_Jg^2m_\pi^2}{48\pi f_\pi^2}\frac{{\rm e}^{-m_\pi r}}{r}\ ,
\end{eqnarray}
from Eq.~(\ref{MDstDstLO}).

\bibliography{reference}

\end{document}